\documentclass[12pt]{article}
\usepackage{epsfig, amssymb}
\newcommand{\be}{\begin{equation}}
\newcommand{\ee}{\end{equation}}
\newcommand{\bea}{\begin{eqnarray}}
\newcommand{\eea}{\end{eqnarray}}
\newcommand{\bA}{\begin{array}}
\newcommand{\eA}{\end{array}}
\newcommand{\bc}{\begin{center}}
\newcommand{\ec}{\end{center}}
\newcommand{\al}{\alpha}

\newcommand{\ra}{\rightarrow}

\newcommand{\p}{\partial}
\newcommand{\ie}{{\it i.e.}}
\newcommand{\eg}{{\it e.g.}}

\newcommand{\Nt}{${\cal N}{=}2$}

\def\BC{{\mathbb C}}
\def\BR{{\mathbb R}}
\def\BZ{{\mathbb Z}}
\def\BM{{\mathbb M}}
\def\BN{{\mathbb N}}

\begin{document}

\begin{titlepage}
%\vspace{10mm}

\bc

\hfill  {Duke-CGTP-04-02} \\
\hfill  {NSF-KITP-04-72} \\
\hfill  {\tt hep-th/0406039} \\
        [22mm]
%\vfill 

{\Huge Localized tachyons in $\BC^3/\BZ_N$}\\ 
\vspace{10mm}

{\large David R.~Morrison, K.~Narayan and M.~Ronen Plesser} \\
\vspace{3mm}
{\small \it Center for Geometry and Theoretical Physics, \\}
{\small \it Duke University, \\}
{\small \it Durham, NC 27708.\\}
\vspace{1mm}
{\small Email : drm, narayan, plesser@cgtp.duke.edu}\\

\ec
\medskip
\vspace{20mm}

\begin{abstract}
We study the condensation of localized closed string tachyons in
$\BC^3/\BZ_N$ nonsupersymmetric noncompact orbifold singularities via
renormalization group flows that preserve supersymmetry in the
worldsheet conformal field theory and their interrelations with the
toric geometry of these orbifolds. We show that for worldsheet
supersymmetric tachyons, the endpoint of tachyon condensation
generically includes ``geometric'' terminal singularities (orbifolds
that do not have any marginal or relevant K\" ahler blowup modes) as
well as singularities in codimension two. Some of the various possible
distinct geometric resolutions are related by flip transitions. For
Type II theories, we show that the residual singularities that arise
under tachyon condensation in various classes of Type II theories also
admit a Type II GSO projection. We further show that Type II orbifolds
entirely devoid of marginal or relevant blowup modes (K\" ahler or
otherwise) cannot exist, which thus implies that the endpoints of
tachyon condensation in Type II theories are always smooth spaces.
\end{abstract}

\end{titlepage}

\newpage 
\begin{tableofcontents}
\end{tableofcontents}

\vspace{5mm}

\section{Introduction}

\par Following the seminal insights of Sen \cite{sen}, the last few years 
have seen the emergence of an understanding of tachyon dynamics in 
string theory.  This development is important both as a foothold on
time evolution in string theory, which has until now been difficult to
study, and because it allows us to study some properties of string
vacua by considering them as the endpoint of tachyon condensation
starting from simpler solutions.  Open string tachyons, being
localized to D-brane  worldvolumes, have relatively controlled
descriptions, obtained by taking limits such that the open string
dynamics decouples from the more complicated dynamics of the bulk
closed-string theory.  In particular, this approach eliminates the
complications of gravitational backreaction.

Closed string tachyons can also exist in configurations in which they
decouple from most of the string modes, and in particular from
gravity.  Consider strings propagating on a space with singularities.  
In an appropriate large radius limit, string modes localized at the
singularity decouple from the bulk theory (a clear general review of
this approach is \cite{emilrev}).  In particular, one can engineer
models in which no tachyons propagate in the bulk while localized
tachyons exist at the singular locus, and study the condensation of
these localized tachyons.  A particularly simple class of
singularities are orbifolds (quotient singularities), 
analyzed in \cite{aps,vafa0111,hkmm} (considerable work has been done 
on closed string tachyon condensation: a recent review with a relatively 
complete list of references is \cite{minwalla0405}). 
In the decoupling limit, the local dynamics near an orbifold point can
be well approximated by the dynamics of string propagation on the
tangent cone $\BC^n/\Gamma$, \ie\ as an orbifold of a free conformal
field theory, and is hence amenable to explicit calculations.  

In fact, what we study is not directly the time evolution of the
system.  Instead, we note that a static tachyon condensate of course
breaks the conformal symmetry.  It corresponds to a relevant
deformation of the worldsheet theory, and we can study the worldsheet
renormalization group flows generated by such deformations.  In
particular, we can consider the endpoints of such flows: as conformal
field theories these correspond to string vacua, and the RG flow thus
realizes a path from one vacuum to another, more stable one.  While
the details of the path are almost certain to differ from the
dynamical time evolution, the endpoints of the flow agree in known
examples with the asymptotic future of the time-dependent solutions.

The free field description of the orbifold theory allows a simple and
direct calculation of the spectrum.  Localized states arise in twisted
sectors, and in appropriate models all tachyons are twisted states.
Unfortunately, this means that in general the free field description 
does not lead to simple descriptions of the deformed theories after
condensation.  If the worldsheet theory with which we start has an
${\cal N}=(2,2)$ superconformal symmetry, there is a class of
deformations for which powerful constraints can be used to control the
RG flow \cite{hkmm}.  These are chiral primary
deformations.\footnote{More precisely, the action is deformed by
adding the integral of the top component of a superfield whose lowest
component is a chiral primary field.} While breaking the conformal
symmetry they preserve the full supersymmetry, simplifying the
description of the deformed models. This simplification is related to 
the existence of a twisted topological version of the theory, retaining 
only the chiral primary fields. 

These decays are also distinguished in that they can be given a clear
geometric interpretation.  They correspond to K\"ahler deformations
(partially) resolving the orbifold singularity, in the sense that the
(non-conformal) supersymmetric field theory obtained by deforming the
action is a nonlinear sigma model on the K\"ahler space obtained by
deforming the quotient. The geometric interpretation provides a 
global setting for the renormalization group flows, and our
understanding of K\"ahler deformation can be used to study global
aspects of the flow.  The description of the resolved space as a toric
variety has proved particularly useful in these investigations \cite{hkmm,
martinecmoore} and will again be useful here.  It turns out that we 
can associate operators directly to particular toric deformations. Taking
the extreme limit of these - when the sizes of all exceptional sets 
are taken to infinity - we find in general several Abelian quotient
singularities in an otherwise smooth space. In general, there will still 
be localized tachyons associated to these, and their condensation will 
continue the process of resolving the singularity.  The geometric 
description, and in particular the description of the resolved orbifold 
as a toric variety, was used in \cite{hkmm} to find the endpoints of the 
decay for cyclic quotient singularities ($\Gamma = \BZ_n$) in complex
codimension one and two. In this paper, we extend this to cyclic 
quotient singularities in codimension three.

\begin{figure}
\bc
\epsfig{file=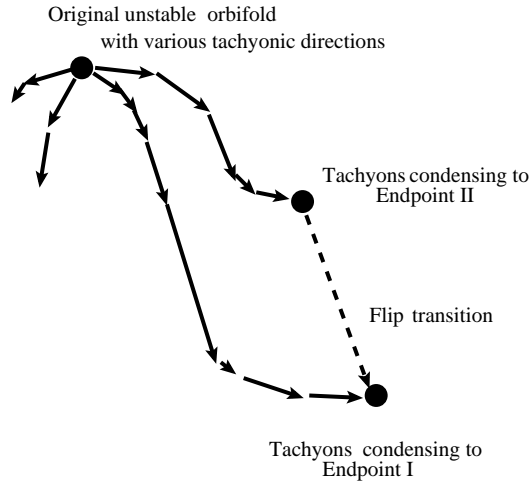, width=7cm}
\caption{A heuristic picture of an unstable (UV) orbifold with several 
relevant directions: two distinct tachyons in $\BC^3/\BZ_{N(p,q)}$ 
condense to distinct endpoints, with a possible flip transition between 
them. Endpoint I stemming from condensation of a more relevant tachyon 
is less singular than Endpoint II (see sec.~4).}
\label{fig7}
\ec
\end{figure}

Some new features of this case are noteworthy. A sequence of
resolutions in the case of codimension two quotients ends either with
a smooth space or with quotients by discrete groups contained in
$SU(2)$.  We will call the latter supersymmetric quotients, because a
type-II string compactification on such a space leads to spacetime
supersymmetry.  A supersymmetric quotient has no relevant operators,
but we can still move out along marginal directions to resolve the
singularity completely. On the other hand, tachyons induce RG flows:
under such a flow to the IR induced by a given tachyon, the anomalous
dimensions of the remaining chiral operators in general increase, as
we show using toric methods. In particular in codimension three, some
operators relevant in the UV become irrelevant in the IR: the blowup
modes corresponding to these modes are no longer tachyonic after the
condensation. In fact, we find situations in which we are left after
condensation with residual quotient singularities for which {\it
all\/} of the chiral blowup modes correspond to {\it irrelevant\/}
operators. As geometric spaces, these are {\it terminal\/}
singularities with no marginal or relevant K\"ahler
blowups,\footnote{Note that by Hironaka's famous theorem on resolution
of singularities \cite{hironaka}, there are always K\"ahler blowup
modes, but most of these are irrelevant \cite{stablesings}.}  well
known in algebraic geometry, and correspond to non-supersymmetric
string vacua with no chiral tachyons: they represent nontrivial
endpoints of the renormalization group flows.  Unlike the case in
codimension one or two, the order in which tachyonic modes condense --
the order in which we blow up -- in general changes the resulting
theory. Thus in general there is no canonical resolution of a given
singularity. Some of the various possible distinct resolutions are
related by {\it flip\/} transitions, similar to the familiar flop
transitions in Calabi--Yau spaces (see figure~\ref{fig7}).

Thinking of renormalization flows provides a natural partial ordering
of the deformation modes of a given quotient: consider a generic
perturbation of the initial theory. Under RG flow, the most relevant
tachyon -- from the point of view of the spacetime theory, this is the
state with the most negative mass squared -- will condense most
rapidly. For some range of initial values, we can obtain an
approximate picture of the actual RG trajectory (our surrogate for the
time evolution) by imagining that we first follow the most relevant
tachyon to a fixed point: in general under the RG flow the original
singularity splits into several singular points\footnote{See \eg\ 
\cite{0308028}, which uses the mirror Landau-Ginzburg description of 
\cite{vafa0111} to show that under condensation of a single tachyon, 
a $\BC^r/\BZ_N$ orbifold decays into $r$ separated orbifolds.}, which 
in the conformal limit, and given the noncompact (large-radius) analysis 
we perform here are decoupled\footnote{Of course this decoupling and
flattening out of the space away from the residual singularities
occurs strictly only in the infinite RG-time limit: during the blowup
process, \ie\ during the condensation of a particular tachyon, the
full space is indeed curved and does not admit any simple free field
theory description.}. Under this flow the conformal weights of other
operators shift, and the most relevant operators remaining will again
be the first to condense in each of the residual singularities.  The
natural prescription is thus to perform the blowup corresponding to
the most relevant tachyon, then repeat the process. This is a partial
ordering because the most relevant field need not be unique, and two
flows corresponding to condensing fields of the same lowest dimension
may lead to different endpoints. It is also worth pointing out that
different decay modes (other than the most-relevant-tachyon sequence)
of the original unstable orbifold are of course possible, giving rise
in principle to distinct geometric endpoints.

Another subtlety in this case is the generic appearance of quotient
singularities in codimension two. These occur along curves contained
within the exceptional sets from the blowups, and the subsequent
twisted sector states that arise are thus in some sense intermediate
between localized and bulk modes. At specific points along the
singular curve the singularity type changes; these quotient
singularities are thus not decoupled even in the extreme infrared
limit, coupled through the twisted modes propagating on the singular
curve.

Following \cite{hkmm} we will use a toric description of the resolved
quotient singularities.  One of the remarkable results of that paper
was the way in which the toric description encodes the algebraic
structure of the chiral ring at the orbifold point.  We find a similar
correspondence in the cases studied here.  Toric descriptions
of chiral rings are known in the case of large-radius limits, but
these are qualitatively different.  The gauged linear sigma model
allows us to interpolate smoothly between the two limits, and the
relation between the two representations is an interesting question,
left for future work. 

Of course, localized tachyons are most interesting in the absence of
bulk tachyons.  As discussed in \cite{atish94, lowestrom, 
aps, vafa0111, hkmm, 0308029} we can, in some
cases, impose a consistent GSO projection on the ${\cal N}=(2,2)$
theory to obtain a modular invariant model from which the bulk tachyon
has been removed, though spacetime supersymmetry is broken. In these
models our goal is to approximate the generic decay.  This, in
general, breaks the worldsheet supersymmetry completely and is thus
not accessible to our methods.  Our approximation consists of a
modification of the procedure mentioned above, of following the most
relevant operator.

The orbifold theory is in fact invariant under three copies of the
${\cal N}=(2,2)$ supersymmetry algebra, and the most relevant operator
is always chiral under some combination of these. We select this
supersymmetry (this choice is closely linked to choosing the target
space complex structure), and use it to follow the renormalization
group trajectory after adding this operator to the action.  At each of
the singularities that remain in the extreme infrared limit, we once
more follow the most relevant operator chiral under the same
supersymmetry.  As a consistency check on the procedure, we show that
the flow does not generate bulk tachyons in various classes of Type II
theories: in other words, all the residual orbifold theories arising
at the ends of our flows are GSO-projected\footnote{After this paper
had been circulated, we learned
that the corresponding analysis in the codimension two case 
which was begun in \cite{0308029} (using
the mirror Landau-Ginzburg
description of \cite{vafa0111}) has been completed in
\cite{0312175}.}.

As above, this process ends when all of the singularities remaining
have no relevant chiral operators.  In fact, since the GSO projection
removes some of the localized tachyons, one finds in general quotients
that are not necessarily {\it terminal\/} singularities in the
geometric sense, but appear {\it string-terminal}, all relevant chiral
operators having been projected out. At this point, though, we once
more have, in each of the decoupled theories, an enhanced
supersymmetry: thus there exist generic metric blowup modes (chiral
with respect to some supersymmetry) that potentially smooth out the
singularities. Indeed, we find a clean combinatoric proof which shows
the non-existence of Type II orbifold singularities completely devoid
of $any$ relevant or marginal blowup modes (K\" ahler or not)
preserved by the GSO projection. Thus not only are there no
string-terminal quotient singularities, there are in fact, for a Type
II string, no terminal singularities at all. This shows that the
endpoints of closed string tachyon condensation for Type II orbifold
string theories in four or more noncompact dimensions are always 
smooth spaces. 

We study the structure of the residual singularities after condensation 
of a single tachyon using toric methods -- in particular, we use the 
``Smith normal form'' of the toric data for the residual geometries to 
glean insight into their structure. 

Organization: we describe the worldsheet conformal field theory of
$\BC^3/\BZ_N$ orbifolds in sec.~2. Sec.~3 describes the representation
via toric geometry of these orbifolds. Sec.~4 follows the renormalization 
group trajectories corresponding to chiral tachyon condensation in Type 0 
string theory: in particular we describe how this dovetails with the 
toric structure of $\BC^3/\BZ_N$ singularities, the analogs of 
``canonical minimal resolutions'' and flip transitions therein, as 
well as the structure of the residual geometries obtained using the 
Smith normal form of the toric data thereof. Sec.~5 describes the 
situation for Type II theories and in particular discusses all-ring 
terminality. Two appendices provide some technical details.

\section{Free field theory at the orbifold point} 

The spectrum of states localized near a quotient singularity is
tractable because, in the limit in which the localized states decouple
from the bulk theory, we are effectively studying the space
$\BC^3/\Gamma$.  As a conformal field theory, this is an orbifold of a
free field theory, obtained by gauging the discrete symmetry group
$\Gamma$.  The high degree of symmetry of the free theory leads to
many simplifications in the treatment of the quotient.

We will work in the RNS formulation, and study the local dynamics near
a singularity of the form $\BR^{1,3}\times\left(\BR^6/\BZ_N\right)$.  
We will choose a complex basis for the ``internal'' coordinates such
that $\Gamma$ acts holomorphically. The generator is thus
\be
\label{gaction}
g: (X_1,X_2,X_3)\to (\omega^{k_1} X_1,\omega^{k_2} X_2,\omega^{k_3} X_3)\ ,
\ee
where $X_i$ are complex coordinates on the internal space and 
$\omega=e^{2\pi i/N}$. 

The free field theory before
orbifolding enjoys an ${\cal N}= (8,8)$ worldsheet supersymmetry,
which will be broken by the quotient (which acts as an $R$-symmetry).
The quotient will preserve three copies of the ${\cal N}=(2,2)$
superconformal algebra, with supercurrents and $U(1)_R$ currents
\bea
\label{ntwo}
G^+_i &=& \psi_i^*\p X_i\nonumber\\
G^-_i &=& \psi_i\p X^*_i\\
J_i &=& \psi_i\psi^*_i = i\, \p H_i\ ,\nonumber
\eea
and their antiholomorphic counterparts, where we have formed complex
linear combinations of the worldsheet fermions as superpartners to the
$X_i$.  We have bosonized the $U(1)$ current so that $\psi_i = e^{i
H_i}$. With respect to this
subalgebra, $(X_i,\psi_i)$ form a chiral superfield, and $g$ acts as a
non-$R$ symmetry with $\psi_i\to \omega^{k_i}\psi_i$. 

Because of the product structure of the free theory, the spectrum of
the quotient theory can be understood by working with one chiral
superfield at a time.  Thus, we consider the theory of one chiral
superfield, and perform a $\BZ_N$ quotient, with the action $X\to
\omega^k X$ (of course, $k$ can be set to one here, but we will want
this peculiar notation later).  Of particular importance to us will be
the ground states in the twisted sectors: twisted-sector states are
the ones that will be localized at the singular locus.  The orbifold
theory will have $N$ twisted sectors and a ``quantum'' $\BZ_N$
symmetry.  In the $j$-th twisted sector, the field $X$ satisfies 
\be
X(\sigma+2\pi,\tau) = \omega^{jk} X(\sigma,\tau)\ .  
\ee 

The ground state in this sector can be shown (see \cite{0308029} for 
a clear exposition) to be a chiral primary state (annihilated by 
$G^+_{-1/2}$ in addition to all positive modes) with $U(1)_R$ charge 
$y=\{ {jk\over N} \}$\ (the fractional part of ${jk\over N}$) and 
conformal weight $h={y\over 2}$, when $y < {1\over 2}$. The first excited 
state is antichiral, with charge $y-1$ and weight $h={(1-y)\over 2}$. 
When $y > {1\over 2}$, the ground state is an antichiral state with 
charge $y-1$ and weight $h={(1-y)\over 2}$, while the first excited state 
is chiral, with charge $y$ and weight $h={y\over 2}$. These exhaust the 
(anti)-chiral states in the theory,  and the results are simply
summarized by the statement that for each $j$ we have a chiral state
of charge $y/2$ and an antichiral state of charge $(1-y)/2$. 

The mass-shell condition gives the mass in spacetime (\ie\ the 
unorbifolded dimensions) of a state with R-charge $q$ and conformal 
weight $h={|q|\over 2}$ as 
\be
-{\al'\over 4} M^2 + h - {1\over 2} = 0 .
\ee
Thus the most relevant tachyon, \ie\ smallest R-charge, corresponds 
to the leading spacetime instability, \ie\ with the most negative 
mass-squared. 

Chiral operators are of interest for several reasons. Since they saturate 
the inequality $h \ge {|q|\over 2}$ between conformal weight and $U(1)_R$ 
charge, their operator products are nonsingular, and by taking the 
coincident limit produce the structure of a ring of (anti-) chiral 
operators. (Keeping track of both in this case is of course a bit 
redundant, since a chiral field in the $j$-th sector has a conjugate 
field in the $(N{-}j)$-th sector.)  Constrained by conservation of 
$U(1)_R$ as well as by the ``quantum'' $\BZ_N$ symmetry of the orbifold 
theory, the structure of the ring is here particularly simple. The 
operator \cite{dfms} creating the chiral state in the $j$-th sector, 
$T_j$, can be written as 
\be\label{onetwistfield}
T_j = \sigma_y e^{iy (H-\overline H)}\ ,
\ee
where $\sigma_y$ is the bosonic twist operator (with conformal weight 
$h={1\over 2}y(1-y)$), and $\p H$ the bosonized current as above. 
The ring is generated by 
\be 
T = \sigma_{1/N} e^{(i/N)(H-\overline H)} \ ,
\ee
with 
\be 
T_j = T^{N y}\ .
\ee

There is, in addition to the identity, one chiral primary operator in
the untwisted sector, 
\be
Y = {1\over V} \psi\overline\psi\ ,
\ee
the volume form of the internal space, normalized by its total
volume.  The two generators satisfy the relation \cite{hkmm}
\be
T^N = Y\ .
\ee
The antichiral ring has a similarly simple structure.  It is helpful 
in the sequel to note that chiral and antichiral fields under the 
algebra (\ref{ntwo}) are exchanged if we exchange $G^\pm$, or 
equivalently ($X$, $\psi$) and ($X^*$, $\psi^*$). 

A chiral (or antichiral) field is the lowest component of an \Nt\ chiral 
superfield whose top component can be added to the action without 
breaking supersymmetry.  This means we can use the powerful constraints 
imposed by $(2,2)$ worldsheet supersymmetry to study the deformed 
theory, \ie\ the RG flow and its endpoints. 
In the string theory, the most relevant operator in any sector is 
chiral (or antichiral), so the tractable sector includes the dominant 
decay modes of these unstable vacua. At any point along the
renormalization group trajectory corresponding to the condensation of
a chiral field, one can perform a topological twist \cite{wittenphases} 
so that the flow corresponds to a family of topological theories (see 
\cite{morrisonplesserInstantons} for the generalization to toric 
varieties). These, in turn, may be amenable to study using a twisted 
gauged linear sigma model (see \eg\ \cite{martinecmoore}). 
In this case, one can follow the condensation of the tachyon all along 
the flow and not simply study its endpoints. 
\\

{\bf \emph{$\BC^3/\BZ_{N(p,q)}$ : chiral rings}}\\
Returning to the case of interest (\ref{gaction}) and recalling that 
the quotient theory is in fact invariant under the three copies of the 
${\cal N}=2$ superconformal symmetry (\ref{ntwo}), we find eight rings 
of operators (anti-) chiral under each of these, in four conjugate 
pairs. In this section, we will focus on one of these, the 
$(c_X,c_Y,c_Z)$ or the chiral ring. Furthermore we will focus 
largely on orbifolds that can be expressed in canonical form, \ie\ 
$(k_1,k_2,k_3)\equiv (1,p,q)$. This includes all isolated orbifolds.\\

We denote noncompact $\BC^3/\BZ_N$ orbifolds with the geometric 
action on the $(X=z^4+iz^5,Y=z^6+iz^7,Z=z^8+iz^9)$ target space coordinates 
\be\label{geomactionC31pq}
(X,Y,Z) \ra (\omega X, \omega^p Y, \omega^q Z), 
\qquad \qquad \qquad |p|,|q|<N,\ \ \ p,q \in \BZ, \ \ \omega=e^{2\pi i/N}
\ee
by $\BC^3/\BZ_{N(p,q)}$. 
String theory on such orbifolds retains no supersymmetry if 
$1+p+q\neq 0({\rm mod} N)$ since the orbifold action does not lie within 
$SU(3)$ -- these orbifolds cannot be embedded as local singularities in 
a Calabi-Yau 3-fold. These are isolated singularities if $p,q$ are 
coprime with respect to $N$. 
The twisted sector operators in the chiral ring of $\BC^3/\BZ_{N(p,q)}$ 
\be\label{XcXcYcZ}
X_j \ = \ \prod_{i=1}^3\ X^{(i)}_{ \{jk_i/N\} } \ = \ 
X^{(1)}_{j/N}\ X^{(2)}_{\{jp/N\} } X^{(3)}_{\{jq/N\} },
\qquad \qquad j=1,2,\ldots N-1
\ee
are constructed out of the twist fields (\ref{onetwistfield}) for each of 
the three complex planes parametrized by $X,Y,Z$. $\{x\}=x-[x]$ denotes 
the fractional part of $x$, with $[x]$ the integer part of $x$ (the 
greatest integer $\leq x$).\footnote{Note that for $m,n>0$, we have 
$\ [{-m\over n}]=-[{m\over n}]-1\ $ and therefore 
$\{ {-m\over n} \}=-{m\over n}-[{-m\over n}]=1-\{ {m\over n} \}$.}
By definition, $0\leq \{x\}<1$. 
This constitutes the $(c_X,c_Y,c_Z)$ ring of twist operators, that are 
chiral with respect to each of the three complex planes. In the $j$-th 
twisted sector, the boundary conditions for the operators $X_j$ are 
\be
X^i(\sigma+2\pi,\tau) = \omega^{jk_i} X(\sigma,\tau)\ . 
\ee
Based on our discussion above in the case of one chiral superfield, the 
chiral operators (\ref{XcXcYcZ}) are either the ground states or the 
first excited states in the various twist sectors.\footnote{For instance, 
in the sector where $\{ {jk_1\over N} \},\{ {jk_2\over N} \}<{1\over 2} ,
\ \{ {jk_3\over N} \}>{1\over 2}$ , the ground state is of the form\ 
$X^{(1)}_{j/N}\ X^{(2)}_{\{jp/N\} } (X^{(3)}_{1-\{jq/N\} })^*$ \ and 
belongs to the $(c_X,c_Y,a_Z)$ ring, which is chiral w.r.t. $X, Y$ and 
anti-chiral w.r.t. $Z$.}

In this notation, we note that the orbifolds $\BC^3/\BZ_{N(p,q)}$, 
$\BC^3/\BZ_{N(-p,q)}$, $\BC^3/\BZ_{N(p,-q)}$ and $\BC^3/\BZ_{N(-p,-q)}$ 
are related by changes of complex structure implemented by the field 
redefinitions $Y\ra Y^*$, $Z\ra Z^*$ and $Y,Z\ra Y^*,Z^*$ respectively. 
As we have seen, besides the $(c_X,c_Y,c_Z)$ ring of operators 
(\ref{XcXcYcZ}), there are various other sets of ``BPS protected'' 
operators which comprise the other rings. It is noteworthy that \eg\ 
the field redefinition $Z\ra Z^*$ exchanges the 
$(c_X,c_Y,c_Z)$ and $(c_X,c_Y,a_Z)$ rings. One can check that the 
$(c_X,c_Y,c_Z)$ ring of the $\BC^3/\BZ_{N(p,q)}$ orbifold is the 
$(c_X,c_Y,a_Z)$ ring of the $\BC^3/\BZ_{N(p,-q)}$ orbifold, \ie\ 
$\BC^3/\BZ_{N(p,N-q)}$ orbifold and similarly for the other rings. 

As a geometric space, by convention the $(c_X,c_Y,c_Z)$ ring of the
singularity alone respects the asymptotic complex structure and
geometry.  In fact, twist operators in the other rings do not appear
as lattice points representing blowup modes of the singularity in the
toric geometric representation of the $(c_X,c_Y,c_Z)$ ring: \ie\ there
is in general a different toric diagram for each of the rings so that
these other rings for a given orbifold do not have an obvious
interpretation in terms of its algebraic geometry. Physically, once a
tachyonic chiral operator condenses, it breaks the full ${\cal
N}=(2,2)$ worldsheet supersymmetry down to the subgroup it preserves,
and the ring it belongs to. Thus if we so wish, we could, for 
noncompact singularities, $define$ the $(c_X,c_Y,c_Z)$ ring to contain 
the most relevant tachyon.
We will have occasion to describe the structure of the twist fields 
in all the various rings later when we discuss Type II theories and 
all-ring terminality. 

Vertex operators belonging to the untwisted sector of these 
$\BC^d/\BZ_N$ orbifolds describe excitations propagating in the full 
ten dimensional spacetime while twisted sector states are localized to 
the singular subspace of the orbifold. The structure of the OPEs of 
general untwisted and twisted sector Virasoro primaries in a regulated 
(noncompact large volume) $V_{10-d}\ra \infty$ limit shows that the 
bulk untwisted sector tachyon of Type 0 decouples and thus can remain 
unexcited along RG flows associated with condensing only the localized 
twisted sector tachyons \cite{hkmm}. Thus it is sensible to study the 
condensation of localized tachyons. 

For the nonchiral Type 0 string theory, one performs a diagonal GSO 
projection which projects out spacetime fermions and retains the bulk 
tachyon. Thus all twisted sector tachyons are present (along with the 
untwisted tachyon). The twist operators $X_j$ have R-charges and 
conformal dimensions 
\be\label{Rh}
R_j \equiv \biggl({j\over N},\biggl\{ {jp \over N} \biggr\},
\biggl\{ {jq \over N} \biggr\} \biggr) 
= {j\over N} + \biggl\{ {jp \over N} \biggr\} 
+ \biggl\{ {jq \over N} \biggr\},
\qquad h_j = {1\over 2} R_j .
\ee
The operators with $R_j<1$ and $R_j=1$ are relevant (tachyonic) and 
marginal respectively while those with $R_j>1$ are irrelevant on the 
worldsheet. In addition to the $X_j$, there are of course the chiral 
primaries, $Y_i={1\over V} \psi_i{\bar \psi_i}$, \ie\ the three volume 
forms of the internal space, normalized by its total volume. There exist 
relations amongst the operators $X_j$, $Y_i$. 

The subset of the twisted states $X_j$ that generates the chiral ring of 
$\BC^3/\BZ_{N(p,q)}$ in general contains more than one element (as in 
the case 
$\BC^2/\BZ_{N(p)}$ studied in \cite{hkmm}). Schematically then a given
operator in the  
chiral ring can be decomposed into products of the generators via the 
ring relation $X_a \sim X_{g_1}^{m_1} X_{g_2}^{m_2} \ldots $, the $m_i$ 
being integers. The R-charge of the generic twisted state $X_a$ is 
given by $R_a = \sum_i m_i R_{g_i}$, where $R_{g_i}$ is the R-charge 
of the generator $X_{g_i}$. A given operator in $\BC^3/\BZ_{N(p,q)}$ 
can be decomposed in various distinct ways so that the generator 
decomposition for a given operator is not unique (as in $\BC^2/\BZ_{N(p)}$). 
This non-uniqueness is fairly obvious from the toric representation 
of these orbifolds, which is 3-dimensional. As we will see, there is 
an intimate relationship between operators in the chiral ring of the 
orbifold and the relations amongst them and the geometry of lattice 
vectors in the toric representation thereof. A noteworthy fact is 
that the set of generators of the $\BC^3/\BZ_{N(p,q)}$ chiral ring in 
general includes irrelevant operators as well as tachyons and marginal 
operators. Indeed as we shall see, there exist classes of 
$\BC^3/\BZ_{N(p,q)}$ orbifolds where the entire chiral ring is 
generated purely by irrelevant operators, \ie\ $R_{g_i}>1$ for all the 
generators $X_{g_i}$. In such cases, there is no relevant or marginal 
deformation of the chiral ring and of the corresponding orbifold 
singularity via K\" ahler blowup modes. 

We will now exhibit some examples elucidating the twisted states $X_j$ 
with their R-charges and the generator decompositions thereof.
\\
{\bf Example $\BC^3/\BZ_{11}\ (1,2,7)$:} The Type 0 theory has tachyons 
$T_1=({1\over 11},{2\over 11},{7\over 11}),\ 
T_2=({2\over 11},{4\over 11},{3\over 11})$ with R-charges 
$\ R_1={10\over 11},\ R_2={9\over 11}\ $ respectively, of which $T_2$ 
survives the chiral GSO projection to Type II. The set of generators 
of the Type 0 chiral ring consists of the tachyonic twist field 
operators $\ X_1=({1\over 11},{2\over 11},{7\over 11}), 
\ X_2=({2\over 11},{4\over 11},{3\over 11})\ $ and the 
irrelevant operators $\ X_5=({5\over 11},{10\over 11},{2\over 11}), 
\ X_6=({6\over 11},{1\over 11},{9\over 11}), 
\ X_7=({7\over 11},{3\over 11},{5\over 11}), 
\ X_8=({8\over 11},{5\over 11},{1\over 11})$. The remainder of the 
twist fields are $\ X_3=({3\over 11},{6\over 11},{10\over 11}), 
\ X_4=({4\over 11},{8\over 11},{6\over 11}), 
\ X_9=({9\over 11},{7\over 11},{8\over 11}), 
\ X_{10}=({10\over 11},{9\over 11},{4\over 11})$. Then it is easy 
to see by comparing R-charges that the relations between various twist 
operators and these generators include 
\be\label{relations1127}
X_3 \sim X_1 X_2, \qquad X_4 \sim X_2^2, 
\qquad X_9 \sim X_7 X_2 \sim X_8 X_1, \qquad 
X_{10} \sim X_8 X_2 
\ee
and other similar expressions. The relation involving $X_9$ illustrates 
the non-uniqueness of the generator decomposition. 
\\
{\bf Example $\BC^3/\BZ_2\ (1,1,1)$:}\ \ The only twisted sector $j=1$ 
of the $(c_X,c_Y,c_Z)$ ring has the irrelevant operator 
$X_1=({1\over 2},{1\over 2},{1\over 2})$ with R-charge 
$R_j={3\over 2}$. Thus the generator of the chiral ring consists of the 
single irrelevant operator $X_1$. It is straightforward to check that 
the twist fields $X_j$ in all rings are irrelevant (as we will see in 
detail later when we discuss Type II theories): This is an isolated 
$all-ring$ terminal singularity. 
\\
{\bf Example $\BC^3/\BZ_N\ (1,p,-p)$:}\ \ ($p,N$ coprime)\ The 
$(c_X,c_Y,c_Z)$ ring twist field $X_j$ has R-charge 
$R_j={j\over N} + \{ {jp \over N} \} + 1 - \{ {jp \over N} \}
=1+{j\over N}>1$ so that these twisted sector states are irrelevant in 
conformal field theory. Thus there are no relevant or marginal operators 
in the $(c_X,c_Y,c_Z)$ ring of the worldsheet string theory describing 
this class of singularities which thus cannot be resolved geometrically 
(see \eg\ \cite{CortiReid}). 
In general however, there are tachyonic or marginal twisted states 
arising from other rings (as we will see in detail later when we discuss 
Type II theories) so that the singularity in general is indeed resolved 
metrically via the nonchiral deformations.

\section{$\BC^3/\BZ_{N(p,q)}$ : toric geometry}

In this section, we will sketch the toric geometry description of 
$\BC^3/\BZ_{N(p,q)}$, uniformizing our notation with the description in 
\cite{hkmm, emilrev} reviewed in Appendix B. The description here 
is based on \cite{psabrgdrm9309, psa9403,
morrisonplesserInstantons, greenerev,
morrisonplesserHorizons} (see \eg\ \cite{fulton} for a detailed 
exposition of toric geometry). 

Let $(x,y,z)$ and $(u,v,w)$ be coordinates on $\BC^3$ and 
$\BC^3/\BZ_{N(p,q)}$ respectively. A basis for the monomials invariant 
under the orbifold action is 
\be\label{basis1}
u=x^N,\ \  v=x^{-p} y,\ \  w=x^{-q} z .
\ee
The ring of holomorphic functions on a neighbourhood of the noncompact 
$\BC^3/\BZ_{N(p,q)}$ orbifold singularity is generated by the monomials 
\be\label{ringbasis}
u^{m_1} v^{m_2} w^{m_3} = x^{Nm_1 - p m_2 - q m_3}\ y^{m_2}\ z^{m_3}
\ee
for integer $m_1,m_2,m_3$. This ring is well-defined if the basis 
functions have positive exponents, \ie\ 
$\ Nm_1-pm_2-qm_3\geq 0, m_2\geq 0, m_3\geq 0$. The space of possible 
such vectors ${\vec m}=\sum_i m_i e_i$ is the cone in the $\BM$ lattice, 
bounded by the vectors $\ e_1=(1,0,0),\ e_2=(p,N,0),\ e_3=(q,0,N)$. 
Thus each point in the $\BM$ lattice defines a monomial on the orbifold. 
The $e_i$ form a basis for the $\BM$ lattice. 

Eqn.~(\ref{basis1}) essentially specializes the general relation 
$\ x_i=\prod_j z_j^{\ a_{ji}},\ i,j=1,2,3\ $ between the coordinates 
$x_i\equiv (u,v,w)\in \BC^3$ and $z_i\equiv (x,y,z)\in \BC^3/\BZ_{N(p,q)}$ 
\ (see eqns.(2.1) and (2.2) of \cite{psa9403}), \ from which we can 
therefore read off the $\ a_{ij}\ $ as 
\be\label{Nvertices}
a_{1j}\equiv \al_1=(N,-p,-q), \qquad a_{2j}\equiv \al_2=(0,1,0), \qquad 
a_{3j}\equiv \al_3=(0,0,1), \qquad j=1,2,3,
\ee 
\ie\ the matrix $a_{ij}$ is formed by juxtaposing the rows 
$\al_1, \al_2, \al_3$. 
Since the orbifold acts as $\ g: z_j \mapsto \omega^{g_j} z_j\ $ 
with rational $g_j$, we have $\sum_j g_j a_{ji} \in \BZ,\ i=1,2,3$. 

The vectors $\al_i$ are constructed orthogonal to the basis $e_i\in \BM$: 
specifically \ $\al_1 \perp e_2, e_3,\ \al_2 \perp e_1, e_3,\ 
\al_3 \perp e_1, e_2$. They form an integral basis for the $\BN$ lattice 
dual to $\BM$. (\ref{Nvertices}) gives the vertices of the simplex 
$\Delta$ defining the fan of cones subtended with the origin $0=(0,0,0)$ 
as the apex in $\BN$. Alternatively one may of course choose to begin 
with a cone in the $\BN$ lattice and construct from it the dual lattice 
$\BM$ as the space of monomials. 

\begin{figure}
\bc
\epsfig{file=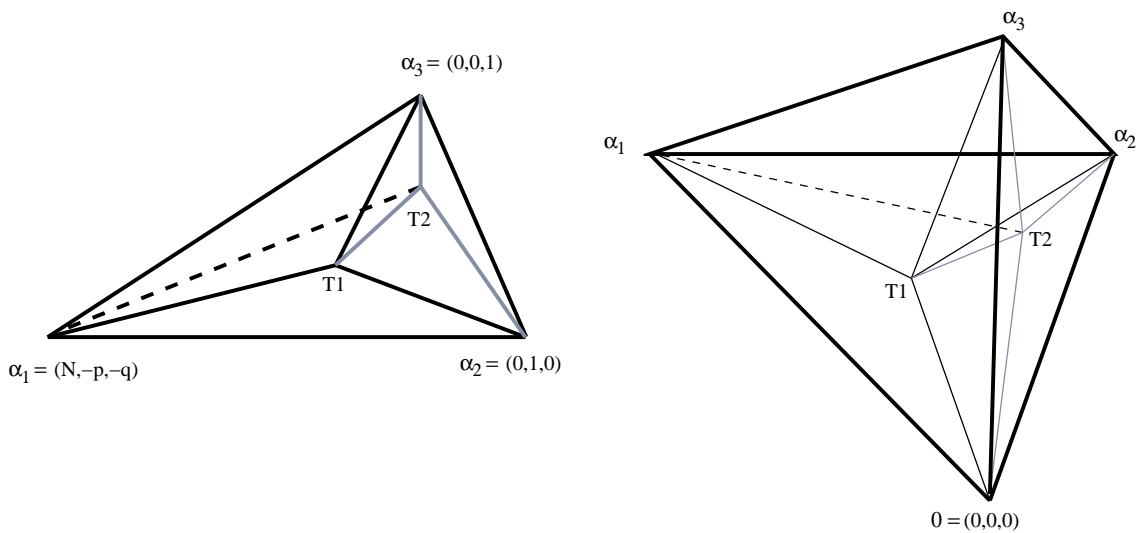, width=15cm}
\caption{The fan of cones for the $\BN$ lattice of $\BC^3/\BZ_{N(p,q)}$, 
with the vertices of the simplex $\Delta$, as well as tachyons $T_1, T_2$ 
in the interior of the cone and the corresponding subdivisions thereof.
(The figure on the left shows the simplex and its subdivision; the
figure on the right shows the actual cones in the fan.)}
\label{fig1}
\ec
\end{figure}
Subcones of the cone $C(0; \al_1,\al_2,\al_3)$ which have (real) dimension 
$p$ determine ``toric''
codimension-$p$ algebraic subspaces of $\BC^3/\BZ_{N(p,q)}$ with (complex) 
codimension $p$.
Thus toric divisors (which are algebraically embedded codimension one 
subspaces) 
are determined by 1-dimensional cones, \ie\ rays in $C(0; \al_1,\al_2,\al_3)$.

Note that the cone $C(0; \al_1,\al_2,\al_3)$
has volume\footnote{The volume of the cube generated by three 3-vectors 
is $V(\al_1,\al_2,\al_3) = 
|{\rm det}(\al_1,\al_2,\al_3)|=|\al_1\cdot \al_2\times \al_3|$.} $N$ 
in terms of the units defined by the lattice $\BN$. 
The relationship between the group action and the lattice $\BN$ in our
case is
visible in the basis in eqn.~(\ref{Nvertices}).
In general, a cone with volume $k>1$ corresponds to an orbifold singularity 
by a discrete group $\Gamma$ whose order is $k$. (If cyclic, this is 
automatically a 
$\BZ_k$ singularity.)  To determine the group $\Gamma$, one needs a
basis for the lattice $\BN$ which is nicely adapted to the group action,
in a manner analagous to eqn.~(\ref{Nvertices}).  In fact, $\BN$ is a
lattice of rank 3 contained in the standard lattice $\BZ^3$, and the
group $\Gamma$ is the quotient $\BZ^3/\BN$, which is a finite group.
The fact that a finitely generated abelian group is a direct sum of
cyclic groups is a standard theorem in algebra (see, for example,
\cite{Lang}), and a specific algorithm to calculate the sum of cyclic
groups -- the ``Smith normal form'' -- will be presented in section 4.

The simplex $\Delta$ is the intersection of the 
fan with an affine hyperplane passing through the three vertices $\al_i$. The 
equation describing the affine hyperplane containing the simplex $\Delta$ 
takes the form $\ell_\Delta(x,y,z)=1$, where
\be
\ell_\Delta(x,y,z) = \biggl( {1+p+q\over N}\biggr) x + y + z.
\ee
We will refer to this affine hyperplane also as $\Delta$. The normal to 
the hyperplane $\Delta$ is the vector $v_{\bot}=(1+p+q,N,N)$, which satisfies 
$\al_i \cdot v_{\bot}=N>0$. The left side of figure~\ref{fig1} shows the 
affine hyperplane 
and the simplex $\Delta$ within it.

There is a remarkable correspondence between operators in the orbifold 
conformal field theory and subspaces in the $\BN$ lattice. 
A given orbifold conformal field theory has relevant/marginal/irrelevant 
operators that correspond to specific lattice points in $\BN$. In our 
normalization, the linear function $\ell_\Delta$ 
evaluated on a specific 
lattice point $j$ 
yields a value
$\Delta_j=\ell_\Delta(x_j,y_j,z_j)$
which
is $equal$ to the R-charge $R_j$ of the corresponding 
operator. A given lattice point $P_j=(x_j,y_j,z_j)$ can then be 
translated to an twisted sector operator as follows: realize that this 
vector can be expressed in the $ \{\al_1,\al_2,\al_3\} $ basis as 
\be
(x_j,y_j,z_j) = a\al_1 + b\al_2 + c\al_3 .
\ee
This then corresponds to an operator $O_j$ with R-charge 
\be
R_j \equiv (a,b,c) = 
\biggl( {x_j\over N},\ y_j+x_j{p\over N},\ z_j+x_j{q\over N} \biggr) .
\ee
Conversely, an operator $O_j$ with R-charge 
$R_j=({j\over N},\{ {jp\over N} \},\{ {jq\over N} \} )$ corresponds to 
a lattice point $P_j=(j,-[{jp\over N}],-[{jq\over N}])$. In general, 
there are lattice points lying ``above'' the affine hyperplane
$\Delta$ that correspond to 
irrelevant deformations. These have $R_j=\Delta_j>1$. 

Conformal field theories corresponding to supersymmetric $\BC^3/\BZ_N$ 
orbifolds always have marginal deformations that are represented by 
points $\beta_k$ that lie $on$ the affine hyperplane
$\Delta$. Thus $R_j=\Delta_j=1$ for the 
$\beta_k$ and we will refer to the affine hyperplane $\Delta$ as the plane of 
marginal operators. The toric picture of blowing up a supersymmetric 
orbifold by a marginal operator then consists of adding to the cone an 
irreducible divisor (\ie\ a ray) corresponding to the marginal operator 
$\beta_k$ and triangulating $\Delta$ into smaller sub-simplices that 
each have $\beta_k$ as a vertex. Such a subdivision of $\Delta$ by one 
or more of the $\beta_k$ gives a subdivision of the cone into subcones, 
corresponding to a blowup that gives a (partial) resolution of the 
orbifold singularity. Subdividing maximally using all the marginal 
operators (\ie\ all the blowup modes) present resolves the orbifold 
completely so that the resulting space is smooth. 
See \eg\ figure~5 
of \cite{psa9403} for a picture of the supersymmetric 
$\BC^3/\BZ_{11}\ (1,3,-4)$ orbifold. In general, there are multiple 
distinct ways to subdivide using the $\beta_k$ in codimension three, 
which give resolved spaces of different topologies: these are 
related by flop transitions. Since the $\beta_k$ lie on $\Delta$, 
the maximal subdivision always yields $N$ minimal subcones each of 
volume one, making up the volume $N$ of the original cone. 

In the nonsupersymmetric cases we study here, in addition to possible 
$\beta_k$ that lie $on$ $\Delta$ corresponding to marginal deformations, 
there are points $T_k$ in the interior of the cone (\ie\ ``below'' 
the affine hyperplane 
$\Delta$) that correspond to relevant (tachyonic) deformations. These 
have $R_j=\Delta_j<1$. 

As we have seen in the previous section, the chiral ring of twist field 
operators is generated by a subset comprising relevant, marginal and 
irrelevant operators. From the point of view of the toric representation 
we have described here, the relations 
$X_a \sim X_{g_1}^{m_1} X_{g_2}^{m_2} \ldots $ between various twist 
operators and the generators (see \eg\ (\ref{relations1127})) are 
precisely equivalent to the different possible ways to generate a given 
lattice vector by adding lattice vectors corresponding to generators. 

The cone corresponding to a nonsupersymmetric singularity can be 
subdivided by the points $T_k$ as well as the $\beta_k$. Figure~\ref{fig1} 
shows tachyons $T_1$ and $T_2$ and their corresponding subdivisions. 
Such a tachyonic subdivision by a $T_k$ results in a partially-resolved 
space with total subcone-volume less than $N$, the volume of the original 
cone: in other words, the resulting space is less singular than the 
original orbifold. In general, if there are multiple tachyonic points 
$T_k$ in the interior, there are multiple distinct ways to subdivide 
which correspond to distinct resolutions of the original singularity, 
typically with distinct total volumes of the subcones. In other words, 
there is in general no canonical resolution. Some of these distinct 
resolutions are related by what are known as \emph{flip transitions}. 
On the other hand, distinct subdivisions which give identical total 
volumes of their corresponding subcones are potentially related by 
flop transitions. In general the subcones obtained after all the 
subdivisions have been executed need not have volume $V_{subcone}=1$ 
each: the endpoint of maximal subdivisions corresponding to a generic 
collection of tachyons in $\BC^3/\BZ_{N(p,q)}$ includes subcones of 
volume $V_{subcone}>1$. 
Such a subcone corresponds to a $geometric$ terminal singularity, 
with no further lattice points on its affine hyperplane $\Delta_{subcone}$ 
or in 
the interior thereof: thus there are no further relevant or marginal 
chiral deformations of the corresponding conformal field theory by 
which it can be resolved via K\" ahler blowups. In the next section, 
we will elaborate more on these phenomena.

\section{Geometric terminal singularities, flip transitions and all that}

Unlike complex codimension one and two orbifold singularities, 
nonsupersymmetric (codimension three) $\BC^3/\BZ_{N(p,q)}$ orbifolds 
generically include ``geometric'' terminal singularities, containing 
no marginal or relevant K\" ahler blowup modes -- a phenomenon well-known 
in algebraic geometry \cite{reid}. In physical terms,
the corresponding worldsheet string conformal field theories do not 
contain any chiral relevant or marginal twisted sector operator by 
which the singularities can be resolved. A simple example is 
$\BC^3/\BZ_2\ (1,1,1)$: the twisted state from the only sector $j=1$ 
has R-charge $R_j={3\over 2}>1$, corresponding to an irrelevant 
operator in conformal field theory. More generally, it is easy to see 
that $\BC^3/\BZ_{N(p,-p)}$ is a geometric terminal singularity for 
$p,N$ coprime: the R-charge of the $j$-th twisted state is 
$R_j=1+{j\over N}>1$, so that the $(c_X,c_Y,c_Z)$ (chiral) twist fields 
in each twisted sector correspond to irrelevant operators in the 
conformal field theory.\footnote{In the supersymmetric $\BC^3/\BZ_N$ 
cases, there always exist marginal chiral deformations which resolve 
the singularity to flat space. On the other hand, in supersymmetric 
$\BC^4/\BZ_N$ (no tachyons) and higher dimensions, there exist 
singularities which cannot be resolved by marginal chiral deformations 
either: the proof of this uses toric techniques combined with results
from Refs.~\cite{rigidity, fine}.}

This result is further strengthened by the fact that toric $\BC^3/\BZ_N$ 
orbifold singularities have no complex structure 
deformations \cite{rigidity}, unlike 
$\BC^2/\BZ_N$: resolutions must be by K\"ahler blowup modes.

On the other hand, sometimes the singularities can indeed be resolved.
A simple example that is devoid of the intricacies to 
follow later is:\\
\begin{figure}
\bc
\epsfig{file=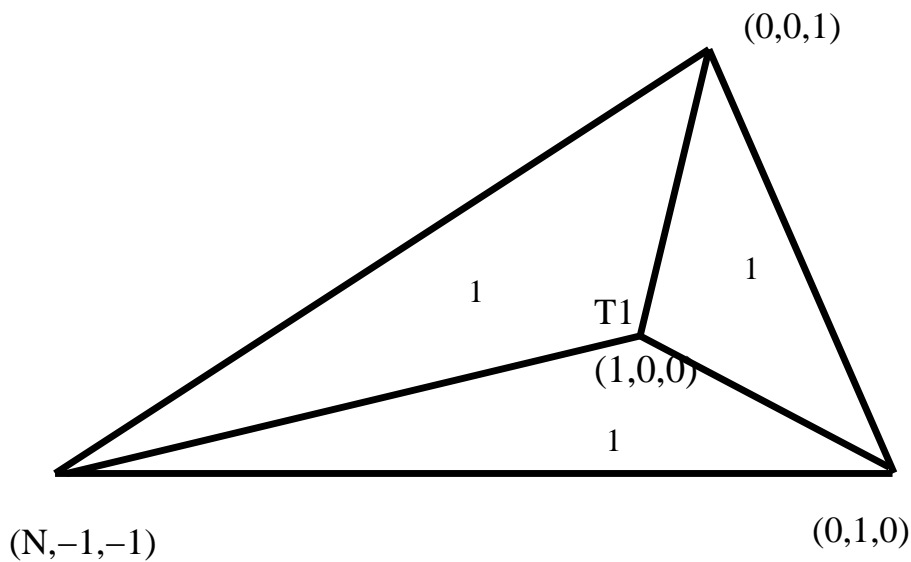, width=12cm}
\caption{$\BC^3/\BZ_N\ (1,1,1)$ : The three points defining the 
affine hyperplane $\Delta$ are shown, along with the sole tachyon $T_1$ and 
its corresponding subdivision.}
\label{fig2}
\ec
\end{figure}
{\bf Example $\BC^3/\BZ_N\ (1,1,1)$:}\ \ Figure~\ref{fig2} shows the 
vertices $(N,-1,-1),(0,1,0),(0,0,1)$ of the cone in the toric diagram. 
For $N=3$, this can be recast as the familiar supersymmetric $\BZ_3$ 
orbifold, while for $N>3$ there is a single tachyonic operator 
$T_1=(1,0,0)$ with R-charge $R_j={3\over N}$ that generates the chiral 
ring (shown in the figure). The operators $T_j=(j,0,0)$ with R-charges 
$R_j={j\over N}$ are tachyonic for $j<N$ and are generated by $T_1$: 
the $T_j$ are collinear with the generator $T_1$ and the origin. As 
can be seen from figure~\ref{fig2}, the three subcones arising from the 
subdivision by $T_1$ have volumes 
$V(0;\al_1,\al_2,T_1)=V(0;\al_2,\al_3,T_1)=V(0;\al_3,\al_1,T_1)=1\ $ 
for any $N$, so that condensation of the most relevant tachyon $T_1$ 
here leads to a resolution of the orbifold to flat space. \\

{\bf \emph{Minimal resolutions, terminal singularities, and flips}}\\
The traditional mathematical approach to studying resolutions of 
singularities (orbifold or otherwise)
proceeds in two steps.  First, thanks to a famous theorem
of Hironaka \cite{hironaka}, it is known that there is a sequence 
of (K\"ahler) blowups which
yields a smooth space.\footnote{Traditionally in algebraic geometry, the
question of K\"ahler metrics on these blowups was not considered, but
since the blowups are projective morphisms, K\"ahler metrics on them
can be chosen so that they indeed become K\"ahler blowups.}
However, this resolution is far from unique, because it is possible
to perform arbitrary additional blowups along smooth subvarieties of
the smooth space, yielding other resolutions. The question thus
arises: can we find some kind of ``minimal'' resolution of a singularity?

In complex dimension one, further blowups do not change the space
and the question does not arise.  In complex dimension two, the
existence of a unique minimal resolution of singularities follows from
the results of the classical Italian school of algebraic geometry
(from the first half of the twentieth century).
The analogous question in complex dimension three was the subject of
intense efforts by mathematicians in the 1980's.

When studying nonlinear sigma models, the possibility of arbitrary blowups
along smooth subvarieties is not an issue for the physical theory, 
because such blowups correspond to irrelevant operators in the
sigma model \cite{stablesings}.  In fact, the classification of
sigma model operators into relevant, marginal, and irrelevant
exactly parallels the criterion introduced by Mori \cite{mori1}
to attack the minimal resolution question: Mori studied the sign of
the intersection numbers $K_X \cdot C$, where $K_X$ is the algebraic
divisor class measuring the zeros and poles of a holomorphic $n$-form,
and $C$ is an arbitrary algebraic curve on $X$.  When $K_X \cdot C<0$,
the sigma model operator which creates $C$ via a blowup
is irrelevant; when $K_X \cdot C = 0 $ the operator is marginal;
and when $K_X\cdot C>0$ the operator is relevant.

Mori showed that starting from a smooth complex threefold, if there are 
any $C$'s with $K_X\cdot C<0$, then there is always a blowdown
map which shrinks some of those $C$'s to zero size.  (This is precisely
what we would expect from analyzing the sigma model, since such
$C$'s correspond to irrelevant operators.)  In complex 
dimension two, the blowdown yields a smooth surface, but in complex
dimension three, the blowdown can introduce singularities, the so-called
{\em terminal singularities}.  By definition (see, e.g., \cite{reid}), 
these are singularities
for which some tensor power $(\Omega^n)^{\otimes r}$ of the
sheaf of holomorphic $n$-forms has local sections,
with the property that for every blowup, the pullback of any local
section of $(\Omega^n)^{\otimes r}$ vanishes along each divisor
created by the blowup. In the case of orbifold singularities, this
rather technical condition can be replaced by a simpler condition
(which is equivalent to checking for relevant or marginal tachyons),
and  the terminal orbifold singularities
in complex dimension three can be completely classified \cite{MS}.

Later study revealed that a minimal resolution could not be reached
by simply following a sequence of blowdowns from the starting resolution:
in addition, one needs to consider ``flips,'' in which (locally) a
single curve $C$ on $X$ shrinks to zero size, but then a different blowup
is done which causes a new curve $C^+$ to grow, creating the space $X^+$.  
We have $K_X\cdot C<0$ but 
$K_{X^+}\cdot C^+ > 0$.  (Both spaces
$X$ and $X^+$ have terminal singularities---there are no flips which
just involve smooth spaces.)  Work by Reid,
Kawamata and others culminated in the final result by Mori \cite{mori2}
which showed that minimal resolutions exist and can be obtained from
arbitrary resolutions by a sequence of blowdowns and flips.

In fact, the theorem had earlier been proven in the case of toric blowups,
blowdowns, and flips by Danilov \cite{danilov}.  The combinatorics of
this process on toric varieties is the tool we are using here.\\

{\bf \emph{Relevance of tachyons, volume minimization and flips}}\\ 
In general, for the orbifolds being studied in this paper,
there are multiple tachyons not collinear with the origin 
in the interior of the cone with distinct R-charges \ie\ order of 
relevance, thus giving rise to distinct subdivisions and potential 
flip (or flop) transitions between the distinct subdivisions. As we 
have mentioned before, the most relevant tachyon (smallest R-charge) 
is the dominant perturbation of the worldsheet theory, and 
correspondingly the dominant unstable direction from the point of view 
of the target spacetime (most negative $M^2$). Thus the most relevant 
tachyon is most likely to condense, followed by the next most relevant 
tachyon and so on. In the $\BC^2/\BZ_{N(p)}$ cases, the Hirzebruch-Jung 
minimal resolution theory ensures that the final endpoint of this 
sequence of most relevant tachyons is a set of decoupled flat space 
regions (see Appendix B). $\BC^3/\BZ_{N(p,q)}$ however has richer 
structure: the generic sequence of tachyonic perturbations leads to 
geometric terminal singularities as we have mentioned above. Furthermore, 
tachyons with different degrees of ``tachyonity'', \ie\ relevance of 
R-charge, give rise with different degrees of likelihood to resolutions 
which generically have distinct topologies. There is an interesting 
calculation that shows the relation between the relevance (R-charges) 
of distinct tachyons and the $\BN$ lattice volumes of the subcones 
resulting from the subdivisions corresponding to those tachyons, \ie\ 
the degree of singularity of the residual geometry. Let us begin with 
the cone $C(0;\al_1,\al_2,\al_3)$ corresponding to the unresolved 
$\BC^3/\BZ_{N(p,q)}$ singularity. Now add a lattice point 
$T_j=(j,-[{j p\over N}],-[{j q\over N}])$ 
corresponding to a tachyon $T_j$. Then the total volume of the three 
residual subcones after subdivision with the tachyon $T_j$ is easily 
calculated: let $V(e_0;e_1,e_2,e_3)$ denote the volume of a cone 
subtended by the vectors $e_1,e_2,e_3$ with the vector $e_0$ being the 
apex. Then (see \eg\ figure~\ref{fig1}) we have 
\bea
V_{subcones}(T_j) &=& V(0;\al_1,\al_2,\al_3) - V(T_j;\al_1,\al_2,\al_3)
\nonumber\\
{} &=& \al_1\cdot (\al_2\times \al_3) - 
\biggl[(\al_1 - T_j)\cdot (\al_2 - T_j)\times (\al_3 - T_j)\biggr]
= N - N(1-R_j) \nonumber\\
{} &=& N R_j < N .
\eea
Now if we consider two tachyons $T_1$ and $T_2$ (as in figure~\ref{fig1}), 
such that $T_1$ is more relevant than $T_2$, \ie\ 
\be\label{T2<T1}
R_1 < R_2, \qquad \qquad 
R_i = {j_i\over N} + \biggl\{ {j_i p \over N} \biggr\} 
+ \biggl\{ {j_i q \over N} \biggr\} .
\ee
This implies 
\be
V_{subcones}(T_1) < V_{subcones}(T_2) .
\ee
This shows that given a singularity, condensation of a more relevant 
tachyon locally leads to a less singular partial resolution. For 
instance, imagine that the less relevant tachyon $T_2$ begins to condense 
(figure~\ref{fig7}): if left unperturbed, the singularity will decay 
and settle down to the resolution with three decoupled sub-singularities 
with total volume $V_{subcones}(T_2)$ of the three corresponding 
subcones. However, in the process of decay, if a small fluctuation 
causes condensation of the more relevant $T_1$, this will dominate over 
the earlier blowup causing a $flip$ transition that dynamically forces 
the singularity to settle down to the less singular resolution with 
volume $V_{subcones}(T_1)<V_{subcones}(T_2)$. From the point of view 
of the toric diagram, condensing the tachyon $T_j$ by blowing up the 
corresponding divisor is associated with adding the corresponding 
lattice point (and the ray thereof) and subdividing thereby. Likewise 
blowing down a divisor corresponds to removing its associated ray and 
the subdivision thereof. In this language, a flip transition can be 
thought of in a manner similar to a flop: as a blowdown combined with 
a blowup, except that in this case there is an associated potential 
hill owing to the fact that relevant operators are at work here (we 
remind the reader that a flop consists of truly marginal deformations 
lying on the moduli space of supersymmetric string vacua). For instance 
in figure~\ref{fig7}, endpoint 1 is less singular than endpoint 2: 
thus there is a potential hill for the flip transition $2 \ra 1$. 

For either partial resolution, the residual subcones generically have 
further tachyons themselves, \ie\ the resulting system is itself unstable 
to tachyon condensation. Thus iterating the above argument sequentially 
suggests that sequential condensation of the most relevant tachyons 
(instead of condensation of generic tachyons) is what leads to a 
decoupled set of residual singularities that is as flat as possible, 
\ie\ with minimal total volume $V_{subcones}$ for the entire set of 
subcones. However the local argument for the inequality above does not 
shed light on the global question of whether sequentially condensing 
the most relevant tachyon indeed leads to the least singular endpoint, 
\ie\ whether the resulting stable singularity (with no further 
tachyons) has minimum total $\BN$ lattice volume $V_{subcones}$. \\

{\bf \emph{The Smith normal form}}\\
Let us now ask what the structure of the residual singularities is. 
For instance (see figure~\ref{fig1}) we obtain the three decoupled 
subcones $C(0;T_1,\al_1,\al_2), C(0;T_1,\al_2,\al_3), 
C(0;T_1,\al_3,\al_1)$ as the endpoint of the condensation of the 
first tachyon, say, $T_1$. Each of these residual subcones represents 
a new orbifold conformal field theory (decoupled from the other 
theories corresponding to the other subcones) which itself generically 
contains further tachyons. In general, such a residual subcone, 
defined by three lattice points, is an orbifold singularity whose 
associated group is determined by the lattice points.  Abstractly, it
is easy to say what the group is:  the three lattice points generate
a subgroup $\BM$ of the standard lattice $\BZ^3$, and the quotient group 
$\BZ^3/\BM$ coincides with 
the orbifold group $\Gamma$.  By a standard theorem in abstract algebra
\cite{Lang},
such a quotient group can always be written as a direct sum of cyclic
groups.

Concretely, the group $\Gamma$ associated with such a 
singularity can be calculated efficiently by an algorithm 
which puts an integer matrix into the
so-called  Smith normal form.  The integer matrix in question describes
the inclusion of $\BM$ into $\BZ^3$ (via a choice of generators), and the 
Smith normal form algorithm produces a new choice of generators which
makes the structure as a direct sum of cyclic groups manifest (as well
as indicating the group action for each of the summands).
This is best illustrated 
by an example. For a given subcone, say $C(0;T_1,\al_2,\al_3)$, 
consider the matrix formed by juxtaposing the column vectors 
$T_1,\al_2,\al_3$ representing the subcone vertices and performing 
the following set of elementary row and column operations on it to 
eventually obtain a diagonal matrix 
\bea\label{smithT23}
\left( \bA{ccc} j_1 & 0 & 0 \\ 
-[{j_1 p\over N}] & 1 & 0 \\
-[{j_1 q\over N}] & 0 & 1 \\
\eA \right)
\ \ra \ 
\left( \bA{ccc} j_1 & 0 & 0 \\ 
-[{j_1 p\over N}] & 1 & 0 \\
0 & 0 & 1 \\
\eA \right)
\ \ra \ 
\left( \bA{ccc} j_1 & 0 & 0 \\ 
0 & 1 & 0 \\
0 & 0 & 1 \\
\eA \right) .
\eea
We have performed the operations $\ T_1+[{j_1 p\over N}]\al_2\ $ and 
$\ T_1+[{j_1 p\over N}]\al_2+[{j_1 q\over N}]\al_3\ $ on the first 
column, in the first and second step here. This then corresponds to 
the singularity $\BZ_{j_1}$, with action 
$(1,[{j_1 p\over N}],[{j_1 q\over N}])$ on the three coordinates 
represented by the vertices $T_1,\al_2,\al_3$.

More generally, begin with the matrix formed by juxtaposing the column 
vectors $v_1,v_2,v_3\in \BZ^3$ representing the vertices of a given 
subcone $C(0;v_1,v_2,v_3)$ and perform elementary row and column 
operations\footnote{These consist of: (1) multiplying a row (column) 
by (-1), (2) adding an integral multiple of one row (column) to 
another, (3) interchanging rows (columns).} to obtain a diagonal 
matrix $\ {\rm diag}[d_1,d_2,d_3]$, with $\ d_3|d_2|d_1$, as the 
Smith normal form of the original matrix: this corresponds to a 
$\BZ_{d_1}\times \BZ_{d_2}\times \BZ_{d_3}$ singularity.\footnote{The fact
that this can always be done is the well-known structure theorem
for finitely presented abelian groups \cite{Lang}.} The Smith 
normal form algorithm\footnote{It is useful to note that the command 
${\bf ismith}$ in Maple performs the Smith algorithm.} essentially 
executes $GL(3,\BZ)$ transformations\footnote{$GL(3,\BZ)$ consists of 
$3\times 3$ matrices with $Det=\pm 1$: the determinant condition is 
equivalent to requiring the entries of the inverse matrix to be 
integral.} on the lattice vectors as well as on the lattice basis 
itself: column operations change the lattice vectors while row 
operations are simply a change of basis of the lattice which thus do 
not affect the relations between the lattice vectors (and therefore 
the structure of the orbifold action). Note that generically both 
column and row operations are required to obtain the Smith normal form, 
putting the lattice vectors describing the orbifold 
into canonical form. As another example, consider 
the following matrix which appears for the subcone $C(0;T_2,\al_1,\al_2)$ 
in $\BC^3/\BZ_{11}\ (1,2,7)$ %(see figure~\ref{fig3}) 
\bea\label{smithal1al2T2}
\left( \bA{ccc} 11 & 0 & 2 \\ 
-2 & 1 & 0 \\
-7 & 0 & -1 \\
\eA \right)
\ \ra \ 
\left( \bA{ccc} 11 & 0 & 2 \\ 
0 & 1 & 0 \\
-7 & 0 & -1 \\
\eA \right)
\ \ra \ 
\left( \bA{ccc} -3 & 0 & 2 \\ 
0 & 1 & 0 \\
0 & 0 & -1 \\
\eA \right) 
\ \ra \ 
\left( \bA{ccc} -3 & 0 & 0 \\ 
0 & 1 & 0 \\
0 & 0 & -1 \\
\eA \right) .
\eea
We have performed the column operations $\ \al_1+2\al_2\ $ and 
$\ \al_1+2\al_2-7T_2\ $ on the first column, in the first and second 
step here, followed by a row operation $r_1+2r_3$ in the third step. 
This thus corresponds to a $\BZ_3\ (-7,1,2)$ singularity (the orbifold 
actions are on $T_2,\al_1,\al_2$ respectively), which is the same as 
$\BZ_3\ (-1,1,2)$. We will find more intricate use for this algorithm 
in what follows. 

There is an alternate equivalent way of realizing the orbifold action, 
which is to find a linear combination of the vertices in question that 
is itself a vector in the lattice generated by the vertices. As an 
example, it is clear from our first example above that the vector 
$\ {1\over j_1}(T_1+[{j_1 p\over N}]\al_2+[{j_1 q\over N}]\al_3)=(1,0,0)\ $ 
is clearly a lattice point itself: from this we read off the orbifold 
action to be $\BZ_{j_1}\ (1,[{j_1 p\over N}],[{j_1 q\over N}])$ as 
above. In the second example above, we have 
$\ -{1\over 3}(-7T_2+\al_1+2\al_2)=(1,0,0)\ $ as a lattice vector, 
giving $\BZ_3\ (-7,1,2)\equiv \BZ_3\ (-1,1,2)$ as above. 

It is important to note that the vector that arises as the linear 
combination in the Smith normal form is only defined up to linear 
combinations of integral multiples of the original lattice vectors.\\

{\bf \emph{Renormalization of subsequent tachyons within subcones}}\\
As mentioned previously, a tachyonic chiral operator that condenses 
breaks the full ${\cal N}=(2,2)$ worldsheet supersymmetry down to the 
subgroup preserved by the ring it belongs to (say $(c_X,c_Y,c_Z)$). 
Furthermore, the residual theories arising at the endpoint of 
condensation of a given tachyon have different R-symmetries from the 
original conformal field theory. Thus the R-charges of the subsequent 
tachyons remaining in the residual geometries get ``renormalized'' 
after a given tachyon has condensed, the specific renormalization of 
a particular subsequent tachyon depending on which of the three 
decoupled subcones it lies within. A subsequent tachyon in a residual 
subcone is a chiral operator with respect to the R-symmetries 
preserved by that subcone. In figure~\ref{fig1} for instance, 
the subsequent tachyon $T_2$ lies in the subcone $C(0;\al_2,\al_3,T_1)$. 
It is then easy to see that the R-charge of $T_2$ is in general 
different, since its geometry relative to the affine hyperplane 
$\Delta_{subcone}$ 
(the plane passing through the subcone vertices $T_1,\al_2,\al_3$) 
is different from before. Since $\Delta_{subcone}$ dips inwards, in 
other words away from the original $\Delta_{cone}$, a subsequent 
tachyon is always closer to $\Delta_{subcone}$ (corresponding to the 
subcone it lies within). Thus the renormalized R-charges of subsequent 
tachyons are always greater than their prior values. 

This renormalization of the R-charge can be calculated by studying the 
geometry of the subsequent tachyon in question relative to the subcone 
it lies within (the expressions appearing below can be checked for 
agreement with $R_k'=a+b+c$ calculated for $T_k$ lying in a subcone in 
the next section on Type II theories). The equation describing an
affine hyperplane 
$\Delta_{subcone}$ passing through three lattice points 
$v_1=(x_1,y_1,z_1), v_2=(x_2,y_2,z_2), v_3=(x_3,y_3,z_3)$ can be 
written as 
\bea
\ell_{\Delta_{subcone}}(x,y,z)\equiv{D(x,y,z)\over D(x_1,y_1,z_1)}=1, \quad 
D(x,y,z)=\det\left( \bA{ccc} x & y & z \\ 
x_2-x_1 & y_2-y_1 & z_2-z_1 \\
x_3-x_1 & y_3-y_1 & z_3-z_1 \\
\eA \right) .
\eea
This has been normalized so that a lattice point lying on 
$\Delta_{subcone}$ corresponds to a marginal operator with R-charge 
$R_j=\ell_{\Delta_{subcone}}(x_j,y_j,z_j)={D(x_j,y_j,z_j)\over D(x_1,y_1,z_1)}=1$. To 
illustrate this expression, consider the subsequent tachyon $T_2$ 
in the residual subcone $C(0;T_1,\al_2,\al_3)$ in figure~\ref{fig1}. 
Then the renormalized R-charge of $T_2$ after $T_1$ has condensed can 
be read off from the equation describing $\Delta_{subcone}$ through 
$T_1,\al_2,\al_3$ (simplifying the above expression using 
(\ref{Rh}) for the R-charge) as 
\be\label{renR1}
R_2'=\ell_{\Delta^{23T}}(T_2)\equiv{D^{23T}(T_2)\over D^{23T}(T_1)}
=R_2 + {j_2\over j_1}\biggl( 1-R_1 \biggr)
=R_2\ \biggl[1 + {j_2\over j_1}\ {(1-R_1)\over R_2}\biggr] .
\ee
Similarly, the corresponding expressions for the renormalized R-charge 
if the subsequent tachyon $T_2$ lies within the residual subcones 
$C(0;T_1,\al_1,\al_2)$ or $C(0;T_1,\al_3,\al_1)$ in figure~\ref{fig1} 
are 
\bea\label{renR2}
&& R_2'=\ell_{\Delta^{12T}}(T_2)\equiv{D^{12T}(T_2)\over D^{12T}(T_1)}
=R_2\ \biggl[1 + { \{ {j_2 q \over N} \}\over \{ {j_1 q \over N} \} }
\ {(1-R_1)\over R_2}\biggr], \qquad \qquad
T_2\in (0;T_1,\al_1,\al_2), \ \ \nonumber\\
{} && R_2'=\ell_{\Delta^{31T}}(T_2)\equiv{D^{31T}(T_2)\over D^{31T}(T_1)}
=R_2\ \biggl[1 + { \{ {j_2 p \over N} \}\over \{ {j_1 p \over N} \} }
\ {(1-R_1)\over R_2}\biggr], \qquad \qquad
T_2\in (0;T_1,\al_3,\al_1) . \nonumber\\
\eea
It is easy to see that since we are considering tachyons, we have 
$R_1<1$ so that $R_2'>R_2$ always. In the examples below, we will 
illustrate this fact with several subsequent tachyons that become 
marginal or near-marginal or irrelevant.

We now present a Type 0 example describing the tachyons therein and the 
toric blowups thereof. It is useful to keep in mind the supersymmetric 
$\BC^3/\BZ_{11}\ (1,3,-4)$ orbifold that is studied in some detail in 
\cite{psa9403}.\\

\begin{figure}
\bc
\epsfig{file=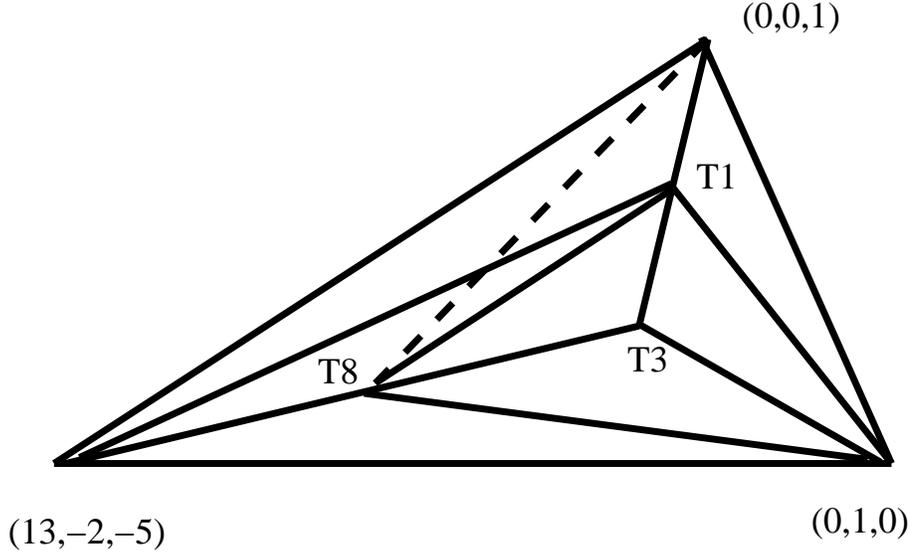, width=12cm}
\caption{$\BC^3/\BZ_{13}\ (1,2,5)$ : The three points defining the 
affine hyperplane $\Delta$ are shown, along with the three tachyons $T_1, 
T_3, T_8$ and two distinct sequences of subdivisions. The solid lines 
correspond to the sequence of most relevant tachyons.}
\label{fig4}
\ec
\end{figure}

{\bf Example $\BC^3/\BZ_{13}\ (1,2,5)$:}\ \ See figure~\ref{fig4}. The 
most relevant tachyon in the Type 0 theory in this example lies in the 
$(c_X,c_Y,c_Z)$ ring and is $T_1=({1\over 13},{2\over 13},{5\over 13})$ 
with R-charge $R_1={8\over 13}$. This ring also has the tachyons 
$T_3=({3\over 13},{6\over 13},{2\over 13})$ and 
$T_8=({8\over 13},{3\over 13},{1\over 13})$ with R-charges \ 
$R_3={11\over 13}$ and $R_8={12\over 13}\ $ respectively. 
In the toric fan (with vertices\ $\al_1=(13,-2,-5),\ \al_2=(0,1,0),\ 
\al_3=(0,0,1)$), these tachyons correspond to the lattice vectors 
$T_1=(1,0,0),\ T_3=(3,0,-1),\ T_8=(8,-1,-3)$. As can be seen from the 
volumes below (or otherwise), $T_1$ and $T_3$ are coplanar with $\al_3$ 
while $T_3$ and $T_8$ are coplanar with $\al_1$. For convenience, we 
note down here the volumes of some subcones that arise in the 
subdivisions we will describe below 
{%\small 
\bea
&& {} V(\al_1,T_3,T_8)=V(\al_3,T_1,T_3)=0,\ \ \ 
V(\al_2,\al_3,T_1)=V(T_3,T_1,T_8)=V(\al_1,T_1,T_8)=1,\ \ \nonumber\\
&& {} V(\al_1,\al_2,T_3)=V(\al_1,\al_3,T_1)=2,\ \ 
V(\al_1,\al_3,T_8)=3,\ \ V(\al_1,\al_2,T_1)=5,\ \ \nonumber\\
&& {} V(\al_3,T_1,T_8)=V(\al_1,\al_2,T_8)=V(\al_2,T_3,T_8)
=V(\al_2,T_1,T_3)=1\ \ 
\eea
}
Let us first consider the subdivisions shown by the solid lines which 
correspond to sequentially blowing up the most relevant tachyon, 
\ie\ the sequence $T_1,\ T_3,\ T_8$: this gives the total volume 
of the subcones to be $6(1)+2$, which is smaller and thus will 
dynamically give rise to a flip transition. Let us consider this 
sequence in more detail: the first subdivision by $T_1$ gives the 
three subcones $C(0;\al_3,T_1,\al_2),\ C(0;\al_1,T_1,\al_3),\ 
C(0;\al_1,\al_2,T_1)$, which by the Smith normal form are respectively 
flat space, $\BZ_2$ and $\BZ_5\ (1,2,-3)$ singularities. Alternatively 
it is straightforward to realize the combinations \ 
${1\over 5} (-\al_1-2\al_2+13T_1) = (0,1,0)$ and \ 
${1\over 2} (-\al_1-5\al_3+13T_1) = (0,1,0)$ as vectors in the lattice: 
these imply that the subcones $C(0;\al_1,\al_2,T_1)$ and 
$C(0;\al_1,T_1,\al_3)$ are respectively 
$\BZ_5\ (1,2,-13) \equiv \BZ_5\ (1,2,-3)$ and $\BZ_2\ (1,5,-13) 
\equiv \BZ_2\ (1,1,-1)$ singularities (shifting by multiples of the 
corresponding lattice vectors). 
Thus the three decoupled residual geometries after this subdivision by 
the most relevant tachyon $T_1$ include a terminal singularity along with 
a supersymmetric orbifold with only marginal blowup modes (\ie\ no 
relevant blowup modes). Using 
(\ref{renR1}) and (\ref{renR2}), we can calculate the renormalized 
R-charges of the subsequent twisted states as 
\be
R_3'=\ell_{\Delta^{12T}}(T_3)={D^{12T}(T_3)\over D^{12T}(T_1)}=1, \qquad 
R_8'=\ell_{\Delta^{12T}}(T_8)={D^{12T}(T_8)\over D^{12T}(T_1)}=1 .
\ee
The R-charges can also equivalently be calculated directly by realizing 
that the corresponding plane is described by $\ x+y+2z=1$ and using 
this to see that the erstwhile tachyons now lie $on$ this plane. 

On the other hand, choosing a different sequence of tachyons by which 
to fully subdivide the cone gives rise to different fans of subcones. 
For instance, the dotted line in figure~\ref{fig4} shows a 
subdivision corresponding to the sequence $T_3,\ T_8,\ T_1$.
From the combinations 
${1\over 3} (\al_3+T_3) = (1,0,0)$, \ 
${1\over 6} (13T_3-2\al_3-3\al_1) = (0,1,0)$ and \ 
${1\over 2} (-3\al_1-6\al_2+13T_3) = (0,0,1)$ of the lattice vectors, 
we see that the subdivision by $T_3$ gives the subcones 
$C(0;\al_2,\al_3,T_3),\ C(0;T_3,\al_3,\al_1),\ 
C(0;\al_1,\al_2,T_3)$, to be $\BZ_3\ (0,1,1),\ 
\BZ_6\ (13,-2,-3)\equiv \BZ_6\ (1,-2,-3)$ and 
$\BZ_2\ (-3,-6,13)\equiv \BZ_2\ (1,0,1)$ singularities respectively. 
Further it is easy to see that\ $T_1={1\over 3} (T_3+\al_3)$ and 
$T_8={1\over 2} (T_3+\al_1)$ : thus $T_8$ becomes marginal after the 
blowup by $T_3$ while $T_1$ acquires the R-charge 
$R_1'={2\over 3}>{8\over 13}$. One can now subdivide by the remaining 
relevant tachyons to obtain the total volume of the subcones to be 
$6(1)+3$ (from the list of volumes above): this is greater than the 
corresponding total volume in the most-relevant-tachyon subdivision, 
as we saw in general earlier.

\section{Type II theories and all-ring terminality} 

As in the Type 0 theory (see Sec.~2), there are eight rings of operators 
(anti-)chiral under each of the three copies of the \Nt\ superconformal 
algebra, in four conjugate pairs. The chiral GSO projection for Type II 
$\BC^d/\BZ_N$ theories preserves spacetime fermions and eliminates the 
bulk tachyon of the untwisted sector: further it acts nontrivially on 
these twist fields representing tachyons localized to the singular 
orbifold subspace, projecting out some of them. We will now describe 
the structure of the various rings in Type II theories in more detail. 
Unlike the supersymmetric $\BC^3/\BZ_{N(p,-p-1)}$ orbifolds where all 
the $R_j=1$ twisted states in the $(c_X,c_Y,c_Z)$ ring, which comprise 
all the blowup modes of the singularity, are preserved and operators in 
the other rings are projected out, the chiral GSO projection for 
nonsupersymmetric orbifolds retains some localized tachyons in each 
ring, as we describe in greater detail below. 

We recall (Appendix A) that for an orbifold $\BC^3/\BZ_N\ (k_1,k_2,k_3)$ 
to allow a Type II chiral GSO projection admitting spacetime fermions, we 
must have $\sum_i k_i = even$. Localized closed string tachyons arise in 
those twisted sectors for which the action on the twist field operators 
is given by 
\be
X_j^r \ra X_j^r \ (-1)^{E_j^r}, 
\ee
where the GSO exponent $E_j^r$ depends nontrivially on the twist sector 
$j$ as well as the specific (anti-)chiral ring $r$ that the twist field 
belongs to. 
From Appendix A.2 (see \eg\ eqns.~(\ref{projccc})~(\ref{projcca})), 
the GSO exponents in Type II theories for twist field operators in the 
various rings of a $\BC^3/\BZ_N\ (k_1,k_2,k_3)$ orbifold are 
\bea
E_j &=& \sum_i \biggl[ {jk_i\over N} \biggr], 
\qquad \qquad \qquad j=1,\ldots,N-1 \nonumber\\
{} &=& {\rm odd}, \qquad \qquad  X_j\ \in\ (c_X,c_Y,c_Z), (c_X,a_Y,a_Z), 
(a_X,c_Y,a_Z), (a_X,a_Y,c_Z) \nonumber\\
{} &=& {\rm even}, \qquad \qquad  X_j\ \in\ (c_X,c_Y,a_Z), (c_X,a_Y,c_Z), 
(a_X,c_Y,c_Z), (a_X,a_Y,a_Z) 
\eea

For example, a $(c_X,c_Y,c_Z)$-ring twist field operator (in the $(-1,-1)$ 
picture) 
\be
X_j = \prod_{i=1}^3\ X^i_{ \{jk_i/N\} } = 
\prod_{i=1}^3\ \sigma_{ \{jk_i/N\} } \ e^{i \{jk_i/N\} (H_i - {\bar H}_i)}, 
\ee
has its conjugate field 
\be
X_j^* = \prod_{i=1}^3\ ( X^i_{ \{jk_i/N\} } )^* 
= \prod_{i=1}^3\ ( X^i_{ \{-(N-j)k_i/N\} } )^* 
= \prod_{i=1}^3\ ( X^i_{ 1-\{(N-j)k_i/N\} } )^*,
\ee
which\footnote{We recall that $\{x\}=x-[x]$ denotes the fractional part 
of $x$, with $[x]$ the integer part of $x$ (the greatest integer 
$\leq x$). By definition, $0\leq \{x\}<1$. Note that, for $m,n > 0$, we 
have $\ [{-m\over n}]=-[{m\over n}]-1\ $ and therefore 
$\{ {-m\over n} \}=-{m\over n}-[{-m\over n}]=1-\{ {m\over n} \}$.} 
clearly lies in the $(N-j)$-th twist sector in the conjugate ring 
$(a_X,a_Y,a_Z)$. Now it is straightforward to see that if $X_j$ is 
preserved, so is its conjugate field in the conjugate ring: 
\bea
E_j &=& \sum_i \biggl[ {jk_i\over N} \biggr] = {\rm odd} \Rightarrow 
\nonumber\\
E_{N-j} &=& \sum_i \biggl[ {(N-j)k_i\over N} \biggr] = 
\sum_i k_i + \sum_i \biggl[ {-jk_i\over N} \biggr] = - E_j - 3 = {\rm even} .
\eea
Similarly for the other rings and their conjugates. 

As we have briefly mentioned in Sec.~2, there is a convenient notation that 
can be used to study and label twist operators in the various rings. To 
illustrate this, note that twist operators in the $(c_X,c_Y,a_Z)$-ring can 
be rewritten as 
\be
X_j^{cca} = X^1_{ \{jk_1/N\} }\ X^2_{ \{jk_2/N\} }\ (X^3_{ 1-\{jk_3/N\} })^* 
= X^1_{ \{jk_1/N\} }\ X^2_{ \{jk_2/N\} }\ (X^3_{ \{-jk_3/N\} })^*, 
\ee
which resemble twist operators in the $(c_X,c_Y,c_Z)$ ring of the orbifold 
$\BC^3/\BZ_N\ (k_1,k_2,-k_3)$ with $X^3 \ra (X^3)^*$ : indeed the R-charges 
of the operators are identical while the condition on their GSO exponents 
\be
E_j = \sum_i \biggl[ {jk_i\over N} \biggr] = even 
\ee
can be re-expressed as 
\bea
E_j^{cca} &=& \biggl[ {jk_1\over N} \biggr] + \biggl[ {jk_2\over N} \biggr] 
+ \biggl[ -{jk_3\over N} \biggr] \nonumber\\
{} &=& E_j + 1 + {\rm even} = {\rm odd},
\eea
so that as expected for a $(c_X,c_Y,c_Z)$ ring operator, the corresponding 
GSO exponent $E_j^{cca}$ is odd. 
As another example, the $(a_X,a_Y,c_Z)$ ring of $\BC^3/\BZ_N\ (k_1,k_2,k_3)$ 
can be expressed as the $(c_X,c_Y,c_Z)$ ring of the orbifold 
$\BC^3/\BZ_N\ (-k_1,-k_2,k_3)$ with $X^1 \ra (X^1)^*, X^2 \ra (X^2)^*$ . 
Generalizing, we see that operators in non-$(c_X,c_Y,c_Z)$ rings of the 
$\BC^3/\BZ_N\ (k_1,k_2,k_3)$ orbifold can be expressed as $(c_X,c_Y,c_Z)$ 
ring operators of a corresponding orbifold with related weights. This 
interrelation between the structure of the various rings can be exploited 
to rewrite the GSO exponents for preserved twist operators in the various 
rings more conveniently, so that they are uniformly $odd$ as expected for 
the re-expressed $(c_X,c_Y,c_Z)$ ring operators. We will find this 
rewriting of the GSO exponents particularly convenient to use when we 
discuss all-ring terminality (see \eg\ eqn.~(\ref{j1E})). 
\\

{\bf \emph{The GSO projection for subcones and residual tachyons 
within}}\\
We now come to the question of studying the geometry of subcones in 
greater detail, in particular in the light of the chiral GSO 
projection for Type II theories\footnote{The question of the Type II 
GSO projection for residual singularities has been studied in the 
codimension two case in \eg\ \cite{0308029, 0312175} via the Landau-Ginzburg 
description of \cite{vafa0111}.}. Consider condensing a GSO-preserved 
twisted state $T_j=(j,-[{j p\over N}],-[{j q\over N}])\equiv
({j\over N}, \{ {jp\over N} \}, \{ {jq\over N} \})$ in a Type II 
$\BC^3/\BZ_{N(p,q)}$ orbifold. Then the chiral GSO projection requires 
that $p+q=odd$ and $[{j p\over N}]+[{j q\over N}]=odd$. This gives 
three subcones $C(0;T_j,\al_2,\al_3), C(0;T_j,\al_1,\al_2), 
C(0;T_j,\al_3,\al_1)$, which are orbifolds of the general form 
$\BC^3/\BZ_n\ (r',p',q')$. In general these are not isolated. As we 
have seen from (\ref{smithT23}), $C(0;T_j,\al_2,\al_3)$ is equivalent 
to the orbifold $\BC^3/\BZ_j\ (1,[{j p\over N}],[{j q\over N}])$, 
the orbifold action being on the coordinates represented by 
$T_j,\al_2,\al_3$ respectively. This clearly admits a Type II GSO 
projection since $r'+p'+q'=1+[{j p\over N}]+[{j q\over N}]=even$. 
The other subcones are somewhat harder to nail down in general but in 
large classes of cases it is possible to glean insight using the toric 
data. 

Consider the subcone $C(0;T_j,\al_3,\al_1)$: using the Smith normal 
form or otherwise, we see that the vector 
\bea\label{smithT31}
(1,0,0) &=& {1\over N \{ {jp\over N} \} } \biggl(
p(T_j + \biggl[{j q\over N}\biggr]\al_3) 
- \biggl[{j p\over N}\biggr](\al_1+q\al_3) \biggr) 
\nonumber\\
{} &=& {1\over N \{ {jp\over N} \} } \biggl(
p T_j - \biggl[{j p\over N}\biggr]\al_1 
+ (p \biggl[{j q\over N}\biggr] - q \biggl[{j p\over N}\biggr]) 
\al_3) \biggr)
\eea
is in the original lattice, showing that the subcone 
$C(0;T_j,\al_3,\al_1)$ is equivalent to the orbifold 
\[\BC^3/\BZ_{N \{ {jp\over N} \} }\ 
(p,-[{j p\over N}],p [{j q\over N}] - q [{j p\over N}]),\]
 the orbifold 
action being on the coordinates represented by $T_j,\al_1,\al_3$ 
respectively. It is important to note that such a linear combination 
of lattice vectors $T_j,\al_1,\al_3$ giving a vector in the original 
lattice is only defined up to adding integer multiples of the lattice 
vectors. Furthermore this vector is degenerate when any of the 
coefficients of $T_j,\al_1,\al_3$ vanishes: in this case, the subcone 
is non-isolated. Similarly, the existence and non-degeneracy of the 
vector 
\be\label{smithT12}
(1,0,0) = {1\over N \{ {jq\over N} \} } \biggl(
q T_j - \biggl[{j q\over N}\biggr]\al_1 + (q \biggl[{j p\over N}\biggr] 
- p \biggl[{j q\over N}\biggr]) 
\al_2) \biggr)
\ee
in the lattice shows that the subcone $C(0;T_j,\al_1,\al_2)$ is 
equivalent to the orbifold
\[\BC^3/\BZ_{N \{ {jq\over N} \} }\ 
(q,-[{j q\over N}],q [{j p\over N}] - p [{j q\over N}]),\]
 the orbifold 
action being on the coordinates represented by $T_j,\al_1,\al_2$ 
respectively.

It is interesting to note, assuming validity of the Smith vectors 
(\ref{smithT31}) (\ref{smithT12}), that these residual orbifold 
singularities admit a Type II GSO projection: \eg\ using the linear 
combination (\ref{smithT31}) for the subcone $C(0;T_j,\al_3,\al_1)$, 
we have 
\bea
r'+p'+q' = p - \biggl[{j p\over N}\biggr] + p \biggl[{j q\over N}\biggr] 
- q \biggl[{j p\over N}\biggr]
&=& p \biggl(1 + odd - \biggl[{j p\over N}\biggr]\biggr) 
- \biggl[{j p\over N}\biggr] - q \biggl[{j p\over N}\biggr]
\nonumber\\
{} &=& -\biggl[{j p\over N}\biggr]\ (1+p+q) + even = even .
\eea
It is important to note that if the order \ $N \{ {jp\over N} \}$ \ of 
the discrete group of the subcone is odd, then a unit multiple of each 
of the lattice vectors $T_j,\al_3,\al_1$ can be added to the vector 
(\ref{smithT31}): in effect, this has shifted the orbifold weights by 
\ $N \{ {jp\over N} \}$, which changes the GSO projection to Type 0 
if \ $N \{ {jp\over N} \} = odd$. 
Similarly, we can use the linear combination (\ref{smithT12}) for the 
subcone $C(0;T_j,\al_1,\al_2)$) to show that this subcone also admits 
a Type II projection. 

Now let us study subsequent twisted states lying in these subcones 
and the GSO projection for them. Consider a sub-twisted state 
$T_k=(k,-[{k p\over N}],-[{k q\over N}])\equiv
({k\over N}, \{ {kp\over N} \}, \{ {kq\over N} \})$ that was preserved 
by the original chiral GSO projection that was performed, \ie\ 
$[{k p\over N}]+[{k q\over N}]=odd$. We can find the renormalized 
R-charge of $T_k$ w.r.t. the R-symmetries of the subcone that it now 
lies in, as follows (see also the corresponding subsection in the 
previous section). If \eg\ \ $T_k\in C(0;T_j,\al_2,\al_3)$, we can 
write $T_k$ as a linear combination of the new lattice basis vectors 
\bea
T_k &=& a T_j + b \al_2 + c \al_3, \nonumber\\
{} {\rm \ie } \qquad 
\biggl(k,-\biggl[{k p\over N}\biggr],-\biggl[{k q\over N}\biggr]\biggr) 
&=& 
a \biggl(j,-\biggl[{j p\over N}\biggr],-\biggl[{j q\over N}\biggr]\biggr) 
+ b(0,1,0) + c(0,0,1),
\eea
which can be solved to give 
\be
R_k'\equiv (a,b,c)= \biggl( {k\over j},\ \ {k\over j} 
\biggl[{jp\over N}\biggr] - \biggl[{kp\over N}\biggr],\ \ 
{k\over j} \biggl[{jq\over N}\biggr] - \biggl[{kq\over N}\biggr] \biggr) .
\ee
Similarly, if $T_k\in C(0;T_j,\al_3,\al_1)$, we can write 
\bea
T_k &=& a T_j + b \al_1 + c \al_3, \nonumber\\
{} {\rm \ie } \qquad 
\biggl(k,-\biggl[{k p\over N}\biggr],-\biggl[{k q\over N}\biggr]\biggr) 
&=& 
a \biggl(j,-\biggl[{j p\over N}\biggr],-\biggl[{j q\over N}\biggr]\biggr) 
+ b(N,-p,-q) + c(0,0,1),
\eea
to get 
\be
R_k'\equiv (a,b,c)= \biggl( { \{ {kp\over N} \}\over \{ {jp\over N} \} }, 
\ \ {k\over N} - {j\over N} { \{ {kp\over N} \}\over \{ {jp\over N} \}}, \ \ 
\biggl\{ {kq\over N} \biggr\} - \biggl\{ {jq\over N} \biggr\} 
{ \{ {kp\over N} \}\over \{ {jp\over N} \} } \biggr) . 
\ee
Likewise if $T_k\in C(0;T_j,\al_1,\al_2)$, we can write 
\bea
T_k &=& a T_j + b \al_1 + c \al_2, \nonumber\\
{} {\rm \ie } \qquad 
\biggl(k,-\biggl[{k p\over N}\biggr],-\biggl[{k q\over N}\biggr]\biggr) 
&=& 
a \biggl(j,-\biggl[{j p\over N}\biggr],-\biggl[{j q\over N}\biggr]\biggr) 
+ b(N,-p,-q) + c(0,1,0),
\eea
to get (essentially $p\leftrightarrow q$) 
\be
R_k'\equiv (a,b,c)= \biggl( { \{ {kq\over N} \}\over \{ {jq\over N} \} }, 
\ \ {k\over N} - {j\over N} { \{ {kq\over N} \}\over \{ {jq\over N} \} }, 
\ \ \biggl\{ {kp\over N} \biggr\} - \biggl\{ {jp\over N} \biggr\} 
{ \{ {kq\over N} \}\over \{ {jq\over N} \} } \biggr) . 
\ee
In all of the above cases, if $0 < a, b, c < 1$, then $T_k$ is an 
interior lattice point in the subcone in question. Furthermore if 
$a+b+c<1$, then $T_k$ is tachyonic. 

Let us now study the exponent required for the GSO w.r.t. the sub-twisted 
state $T_k$. From Appendix A we recall that for a twisted state $T_k$ 
with R-charge $R_k=(r_1,r_2,r_3)$ in a $\BC^3/\BZ_n\ (k_1,k_2,k_3)$ 
orbifold, the Type II GSO exponent (\ref{gsorpq}) in question can be 
written in terms of the H-shifts $\ a_i=odd$ integers as 
\be
E = \sum_i a_i r_i, \qquad {\rm with}\ \ a_i={\rm odd,\ satisfying}\ 
\sum_i a_i k_i = 0\ ({\rm mod} 2n) .
\ee
Since the original cone is a Type II orbifold singularity, we have 
$p+q=odd$: furthermore, since $T_j$ is preserved, 
$[{j p\over N}]+[{j q\over N}]=odd$. Now consider 
$T_k\in C(0;T_j,\al_2,\al_3)\equiv 
\BC^3/\BZ_j\ (1,[{j p\over N}],[{j q\over N}])$. Then we require $a_i=odd$ 
satisfying $a_1 + a_2 [{j p\over N}] + a_3 [{j q\over N}]= 0\ (mod\ 2j)$: 
this is solved by $a_1=[{j p\over N}]+[{j q\over N}], a_2=a_3=-1$, each 
of which is odd. Then we have 
\be
E = \biggl(\biggl[{j p\over N}\biggr]+\biggl[{j q\over N}\biggr]\biggr) 
{k\over j} - (1) \biggl( 
{k\over j} \biggl[{jp\over N}\biggr] - \biggl[{kp\over N}\biggr] \biggr)
- (1) \biggl( 
{k\over j} \biggl[{jq\over N}\biggr] - \biggl[{kq\over N}\biggr] \biggr)
= \biggl[{kp\over N}\biggr]+\biggl[{kq\over N}\biggr], 
\ee
which is the same as the original GSO exponent for the state $T_k$. 
Thus a sub-twisted state that was preserved by the original GSO, \ie\ 
$E=[{kp\over N}]+[{kq\over N}]=odd$ remains preserved after the 
subdivision by $T_j$. Similarly if $T_k\in C(0;T_j,\al_3,\al_1)\equiv 
\BC^3/\BZ_{N \{ {jp\over N} \} }\ 
(p,-[{j p\over N}],p [{j q\over N}] - q [{j p\over N}])$, we require 
$a_i=odd$ satisfying 
$a_1 p - a_2 [{j p\over N}] + a_3 p [{j q\over N}] - a_3 q [{j p\over N}] 
= 0\ (mod 2N \{ {jp\over N} \})$: 
this is solved by $a_1=[{j p\over N}]-[{j q\over N}], a_2=p-q, a_3=1$, 
each of which is odd. Then we have 
\be
E = \biggl( \biggl[{kp\over N}\biggr]+\biggl[{kq\over N}\biggr] \biggr)
+ even, 
\ee
and likewise for the case $T_k\in C(0;T_j,\al_1,\al_2)\equiv 
\BC^3/\BZ_{N \{ {jq\over N} \} }\ 
(q,-[{j q\over N}],q [{j p\over N}] - p [{j q\over N}])$. 

Thus this proves that originally preserved sub-twisted states continue to 
be preserved after condensation of a preserved twisted state for each of 
the three subcones. It is important to note however that our proof really 
only holds for the cases where the subcone in question is in fact the 
orbifold determined by Smith normal form as above. Indeed as we will see 
in the examples to follow, condensation of a tachyon in an isolated 
orbifold does give rise to residual non-isolated singularities with 
some sub-twisted states lying on a wall of one or more of the subcones. 
In such cases, the above expression of the Smith normal form that we 
have used could potentially break down, although in specific examples, 
it is straightforward to analyze the singularity directly. 
On this note, it is important to realize that an orbifold 
$\BC^3/\BZ_N\ (k_1,k_2,k_3)$ can be written in canonical form 
$\BC^3/\BZ_N\ (1,p,q)$, only if at least one of the $k_i$ is relatively 
prime to $N$, \ie\ $gcd (k_i,N)=1$ for some $k_i$, in which case the 
orbifold has a chance of being isolated. As an example, the non-isolated 
orbifold $\BC^3/\BZ_6\ (2,2,3)$, which cannot be expressed in canonical 
form, cannot be a Type II orbifold since $\sum_i k_i=7=odd$. Although 
the proof above does not shed general light on whether an orbifold such 
as this can arise under condensation of a localized tachyon in a Type II 
orbifold, this seems unlikely: in particular, we have not found any 
Type II example where a residual singularity does not admit any Type II 
GSO projection. \\

{\bf \emph{All-ring terminality}}\\
Given the structure of the chiral GSO for the various rings that we have 
described above, it is interesting to ask if the chiral GSO projection 
allows geometric terminal singularities to exist as physically sensible 
Type II theories that admit spacetime fermions projecting out the bulk 
tachyon. As we have seen, the orbifold $\BC^3/\BZ_N\ (1,p,N-p)$ with 
$N=odd$ and $p, N$ coprime is a geometric terminal singularity with all 
twisted states in the $(c_X,c_Y,c_Z)$ ring being irrelevant (see the 
example in Sec.~2): this can also admit propagation of Type II strings, 
since $\sum k_i=1+N=even$ implies that the bulk tachyon can be GSO 
projected out. 

However, this does not preclude the existence of GSO-preserved tachyonic 
twisted states in the other rings. Indeed while these will not be chiral 
deformations of the theory, they are nonetheless blowup modes which 
metrically smooth out the singularity. Thus in order to study the 
physical existence of terminal singularities, we must look for preserved 
relevant operators arising from any of the various rings. 

To study all-ring terminality, it is sufficient for convenience to
label distinct orbifolds of the form $\BZ_N\ (1,p,q)$ by restricting
to the case $p,q > 0$.  In this notation, the $(c_X,c_Y,c_Z)$ ring of
the familiar supersymmetric orbifold is recognized as that ring of the
orbifold $\BC^3/\BZ_N\ (1,p,q)$ such that $1+p-q=0\ (mod\ N)$, or
$1-p+q=0\ (mod\ N)$.

Then for the $j=1$ sector in some ring to be preserved, we require the 
corresponding GSO exponent 
\bea\label{j1E}
E_{ccc} &=& \biggl[ {jp\over N} \biggr] + \biggl[ {jq\over N} \biggr] = 0, 
\qquad \qquad 
E_{cca} = \biggl[ {jp\over N} \biggr] + \biggl[ -{jq\over N} \biggr] = -1, 
\nonumber\\
{} 
E_{cac} &=& \biggl[ -{jp\over N} \biggr] + \biggl[ {jq\over N} \biggr] = -1, 
\qquad \qquad 
E_{caa} = \biggl[ -{jp\over N} \biggr] + \biggl[ -{jq\over N} \biggr] = -2, 
\eea
to be odd. From the above, we see that the $j=1$ state survives in the 
$(c_X,c_Y,a_Z)$ and $(c_X,a_Y,c_Z)$ rings. These have the R-charges 
$({1\over N}, {p\over N}, 1-{q\over N})$ and 
$({1\over N}, 1-{p\over N}, {q\over N})$. For the orbifold to be 
``string-terminal'', it is necessary (but not sufficient) to have 
these surviving $j=1$ states be irrelevant: 
\be
{1\over N} + {p\over N} + 1 - {q\over N} > 1, \qquad \qquad 
{1\over N} + 1 - {p\over N} + {q\over N} > 1, 
\ee
\ie\ we want $1+p > q,\ 1+q > p$. This gives $p > q-1$. 

Now recall that to admit a Type II GSO projection, we require $p+q=odd$. 
This implies that $p \neq q$ (since, as we have mentioned above, including 
all rings of the theory, it suffices\footnote{Note that, \eg\ shifting one 
of the orbifold weights by $N=odd$, is already taken care of by including 
all rings.} to consider $p,q > 0$). We can assume without loss of 
generality that $p < q$. This gives 
\be
q-1 < p < q,
\ee
which has no solution for integer $p,q > 0$. 

This shows that Type II all-ring terminality is not possible whenever
the orbifold action is written in the canonical form $(1,p,q)$ : this
includes all isolated orbifolds. However, as we have mentioned in the
previous subsection, the orbifold action $(k_1,k_2,k_3)$ for
non-isolated singularities (as before, restrict to $k_i > 0$, since
all rings are being considered) can be cast in the form $(1,p,q)$,
only if at least one of the $k_i$ is relatively prime to $N$, \ie\
there exists some $l$ such that $\{ {l k_i\over N} \} = 1$ for some 
$k_i$. Note however that for a strongly non-isolated singularity, 
there is no $k_i$ coprime with $N$, \ie\ there always exist some 
$l_i$ such that $\{ {l_i k_i\over N} \} = 0$, for $i=1,2,3$ : such 
twist sectors are likely to be the minimum R-charge sectors, \ie\ 
the sectors where some ring is likely to have a tachyon. Therefore, 
without loss of generality, assume $\{ {l_1 k_1\over N} \} = 0$. 
Then the R-charges for the corresponding states from the 
$(c_X,c_Y,c_Z), (c_X,c_Y,a_Z), (c_X,a_Y,c_Z), (c_X,a_Y,a_Z)$ rings are 
\bea
&& R_l^{ccc} = \biggl\{ {l k_2\over N} \biggr\} 
 + \biggl\{ {l k_3\over N} \biggr\}, \nonumber\\
&& R_l^{cca} = \biggl\{ {l k_2\over N} \biggr\} + 1 - 
\biggl\{ {l k_3\over N} \biggr\}, 
\nonumber\\
&& R_l^{cac} = 1 - \biggl\{ {l k_2\over N} \biggr\} + 
\biggl\{ {l k_3\over N} \biggr\}, \\
&& R_l^{caa} = -\biggl\{ {l k_2\over N} \biggr\} - 
\biggl\{ {l k_3\over N} \biggr\} + 2 , \nonumber
\eea
with the corresponding GSO exponents being $E_l$ for the 
$(c_X,c_Y,c_Z), (c_X,a_Y,a_Z)$ states and $E_l+1$ for the 
$(c_X,c_Y,a_Z), (c_X,a_Y,c_Z)$ states (upto even integers, which do 
not affect the sign of the exponent). Notice that these look like 
the corresponding expressions for codimension two. 
Now if $E_l = even$, the $(c_X,c_Y,c_Z), (c_X,a_Y,a_Z)$ states are 
projected out, and we require for string-terminality that the 
GSO-surviving states are irrelevant, in other words, 
$R_l^{cca}, R_l^{cac} > 1$, \ie\ 
\be
\biggl\{ {l k_2\over N} \biggr\} > \biggl\{ {l k_3\over N} \biggr\} , 
\qquad 
\biggl\{ {l k_3\over N} \biggr\} > \biggl\{ {l k_2\over N} \biggr\} ,
\ee
which strict inequality has no solution. Similarly, if $E_l = odd$, 
then we require for string-terminality that $R_l^{ccc}, R_l^{caa} > 1$ :
again it is straightforward to show the absence of any solution. 
Similarly if $\{ {l_i k_i\over N} \} = 0$, for $i=2,3$. Equality 
in these expressions indicates the appearance of marginal states, which 
can of course resolve the singularity. This shows that there are always 
tachyonic or marginal operators that arise in such sectors. 

This negative result for string-terminality shows that the endpoints
of condensation of all tachyons in Type II nonsupersymmetric unstable
$\BC^3/\BZ_N$ orbifolds are always smooth spaces. In other words, Type
II string propagation in four noncompact dimensions always resolves
potential orbifold terminal singularities.

Along these lines, it is also interesting to ask if there are all-ring
terminal $\BC^3/\BZ_N\ (1,p,q)$ singularities in the Type 0 theory. In
this case, all twisted states in all rings are preserved by the
diagonal GSO projection.  Then consider the $j=1$ sector as before. It
is necessary that the $j=1$ states be irrelevant for terminality. We
therefore require that the R-charges for the corresponding states in
the various rings satisfy
\bea
&& {1\over N} + {p\over N} + {q\over N} > 1, \qquad \qquad 
{1\over N} + {p\over N} + 1 - {q\over N} > 1, \nonumber\\
{} && {1\over N} + 1 - {p\over N} + {q\over N} > 1, \qquad \qquad 
{1\over N} + 1 - {p\over N} + 1 - {q\over N} > 1 .
\eea
These simplify to give 
\be
N-1 < p+q < N+1, \qquad \qquad 
1+p > q, \qquad  1+q > p .
\ee
Now if $p < q$, then as before we have $q-1 < p < q$, which is not possible 
for integer $p,q > 0$. Therefore consider $q=p$. Then we have 
$N-1 < 2p < N+1$. If $N=odd=2l+1,$ we have $2l < 2p < 2l+2$, which is not 
possible for integer $p,l > 0$. If $N=even=2l$, we have $2l-1 < 2p < 2l+1$, 
which gives $p=l={N\over 2}$. This is the orbifold
$\BC^3/\BZ_N\ (1, {N\over 2}, {N\over 2})$. In this case however, the 
$j=2$ twisted state, if it exists, has R-charges $({2\over N}, 0, 0)$, 
which is irrelevant only if $N < 2$ : this however does not give any 
nontrivial orbifold. Therefore the $j=2$ state does not exist, \ie\ $N=2$. 
Thus the only isolated all-ring terminal singularity is 
$\BC^3/\BZ_2\ (1,1,1)$: the only twisted states, coming from the $j=1$ 
sector, are irrelevant in all rings since the corresponding R-charges 
satisfy 
\bea
&& (c_X,c_Y,c_Z): \qquad {1\over 2} + {1\over 2} + {1\over 2} > 1, 
\nonumber\\
{} && (c_X,c_Y,a_Z): \qquad {1\over 2} + {1\over 2} + 1 - {1\over 2} > 1,
\nonumber\\
{} && (c_X,a_Y,c_Z): \qquad {1\over 2} + 1 - {1\over 2} + {1\over 2} > 1, 
\\ {} && 
(c_X,a_Y,a_Z): \qquad {1\over 2} + 1 - {1\over 2} + 1 - {1\over 2} > 1 
\nonumber .
\eea
Note that $\BC^3/\BZ_2\ (1,1,1)$ does not admit a chiral GSO projection 
since $\sum k_i = odd$ : thus it does not admit propagation of Type II 
strings. 

Given all this, we expect the following: physically in a given unstable 
(UV) orbifold, the most relevant tachyon(s) will belong to one (or more) 
of the (anti-)chiral rings. When this condenses, it creates an expanding 
bubble of flat space blowing up the divisor it corresponds to. Metrically 
we expect that this most relevant tachyon pulse will trigger condensation 
of all tachyons within the same ring, since they preserve the same 
fraction of the original ${\cal N}=(2,2)$ supersymmetry. The endpoint 
of condensation of all tachyons within this ring will in general include 
$geometric$ terminal singularities. However as the nonexistence proof 
above shows, these will contain further $nonchiral$ blowup modes which 
will metrically smooth out the corresponding singularities. 

We describe some Type II examples below. 
\\
{\bf Example $\BC^3/\BZ_{13}\ (1,2,5)$:}\ \ Recall the Type 0 example 
we discussed earlier (figure~\ref{fig4}): the most relevant tachyon 
$T_1^{ccc}\equiv({1\over 13},{2\over 13},{5\over 13})={8\over 13}$ 
belonged to the $(c_X,c_Y,c_Z)$ ring. We saw there that the endpoint 
of the most relevant tachyon sequence included the all-ring terminal 
singularity $\BC^3/\BZ_2\ (1,1,1)$. As we have seen above, this does 
not admit a Type II GSO projection so that tachyon condensation 
in this Type 0 theory cannot result in a Type II theory.

\begin{figure}
\bc
\epsfig{file=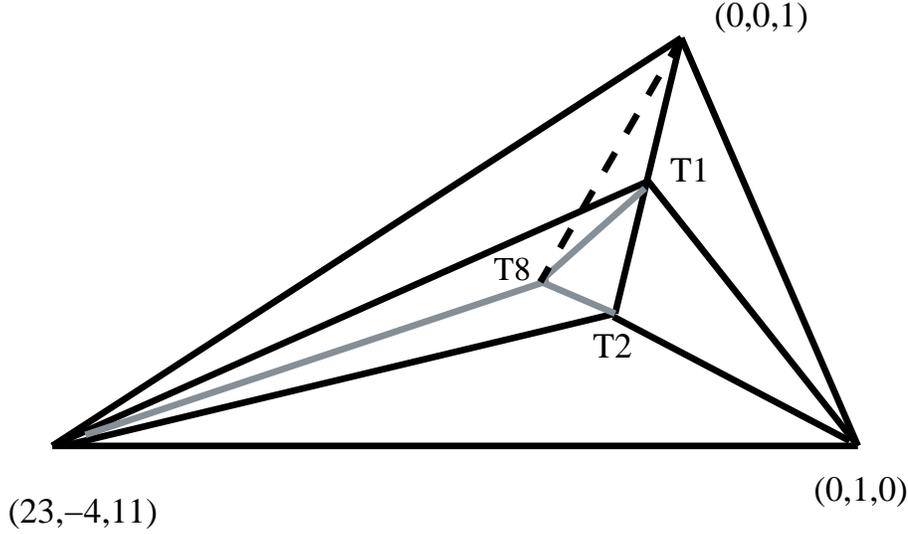, width=12cm}
\caption{$\BC^3/\BZ_{23}\ (1,4,-11)$ : The three points $\al_1, \al_2, 
\al_3$ defining the affine hyperplane $\Delta$ are shown, along with the 
five tachyons $T_2, T_1, T_8$ in Type II. The solid lines show the 
subdivisions corresponding to the sequence of most relevant tachyons 
while the dotted lines show some possible flips.}
\label{fig6}
\ec
\end{figure}

On the other hand, $\BC^3/\BZ_{13}\ (1,2,5)$ itself admits a consistent 
Type II GSO projection. In this case, it is straightforward to see that 
$T_1^{ccc}$ above is in fact GSO-projected out. Of the GSO-surviving 
tachyons, there turn out to be $two$ distinct tachyons with the 
same R-charge $R={9\over 13}$ : these are \ 
$T_2^{cca}\equiv ({2\over 13},{4\over 13},{3\over 13})$ \ and \ 
$T_5^{caa}\equiv ({5\over 13},{3\over 13},{1\over 13})$, \ belonging to 
the $(c_X,c_Y,a_Z)$ and $(c_X,a_Y,a_Z)$ rings respectively. The 
subdivisions thereof and the endpoints of tachyon condensation for 
either of them condensing alone are straightforward to work out. On the 
other hand, the methods we use fail if both condense simultaneously: 
condensation of such mixed tachyons breaks ${\cal N}=(2,2)$ worldsheet 
supersymmetry. 
\\
{\bf Example $\BC^3/\BZ_{23}\ (1,4,-11)$:}\ \ See figure~\ref{fig6}. 
We consider the Type II theory here: the $(c_X,c_Y,c_Z)$ ring tachyons 
$T_1=({1\over 23},{4\over 23},{12\over 23}),\ 
T_2=({2\over 23},{8\over 23},{1\over 23}),\ 
T_8=({8\over 23},{9\over 23},{4\over 23}),\ $ with R-charges 
$R_1={17\over 23},\ R_2={11\over 23},\ R_8={21\over 23}\ $ respectively 
survive the chiral GSO projection. While there are GSO-preserved tachyons 
in the other rings, the most relevant tachyon in this theory in fact is 
$T_1$ above, from the $(c_X,c_Y,c_Z)$ ring.\footnote{One could, if one 
so wishes, $define$ the orbifold action so that the most relevant tachyon 
lies in the $(c_X,c_Y,c_Z)$ ring.}
The vertices of the affine hyperplane of marginal operators are 
$\al_1=(23,-4,11),\ \al_2=(0,1,0),\ \al_3=(0,0,1)$ while the tachyons 
correspond to the lattice vectors 
$T_1=(1,0,1),\ T_2=(2,0,1),\ T_8=(8,-1,4)$. \ $T_1$ and $T_2$ are 
coplanar with $\al_3$. The volumes of some subcones are 
{%\small 
\bea
&& {} V(\al_1,\al_2,T_8)=V(\al_1,T_2,T_1)=V(\al_3,\al_1,T_1)=4,\ \ 
V(\al_2,\al_3,T_2)=2,\ \ V(\al_1,T_2,T_8)=1, 
\nonumber\\
&& {} V(\al_3,T_1,T_8)=V(\al_2,\al_3,T_1)=V(\al_2,T_1,T_2)=
V(T_1,T_2,T_8)=V(\al_1,\al_2,T_2)=1, \ 
\nonumber\\
&& {} 
V(\al_2,T_2,T_8)=0,\ \ V(\al_2,\al_3,T_8)=V(\al_3,\al_1,T_2)=8,\ \ 
V(\al_3,\al_1,T_8)=9. 
\eea
}
As before let us analyze the sequence of most relevant tachyons, \ie\ 
$T_2, T_1, T_8$. Condensation of $T_2$ gives the residual subcones 
$C(0;\al_1,\al_2,T_2),\ C(0;T_2,\al_2,\al_3),\ C(0;T_2,\al_3,\al_1)$,\ 
which correspond to flat space, $\BZ_2\ (1,0,-1)$ and 
$\BZ_8\ (23,-1,-2) \equiv \BZ_8\ (1,1,2)$ singularities respectively, 
using the Smith normal form: alternatively this can be seen by realizing 
the combinations \ ${1\over 2} (T_2-\al_3) = (1,0,0)$ and \ 
${1\over 8} (23T_2-\al_3-2\al_1) = (0,1,0)$ of the lattice vectors 
(shifting by integer multiples thereof). 
Since $\ T_1={1\over 2}(T_2+\al_3)$, it is clear that $T_1$ is marginal 
after condensation of $T_2$. It is straightforward to work out the 
twisted states of these residual orbifolds and map $T_1$ onto the 
corresponding new twisted sector (it is important however to be careful 
in finding the correct Type II projection for the residual orbifolds 
which is consistent with the original theory). On the other hand, the 
subsequent tachyon $T_8$ with renormalized R-charge 
\be
R_8'=\ell_{\Delta^{31T}}(T_8)={D^{31T}(T_8)\over D^{31T}(T_2)}=
{21\over 23}\ \biggl[\ 1 + {9\over 8}\ {1-{11\over 23}\over {21\over 23}}\ 
\biggr] = {3\over 2} > 1
\ee
has become irrelevant after $T_2$ condenses! In fact, we have \ 
$T_8 = {9\over 8}T_2+{1\over 4}\al_1+{1\over 8}\al_3$, \ \ie\ one of 
the coefficients is greater than unity. In figure~\ref{fig6}, 
the solid lines correspond to the sequence of most relevant tachyons, 
while the lightly shaded lines correspond to the subdivision by the 
now irrelevant $T_8$. The total volume of the subcones with this 
sequence of subdivisions is $V_{total}=8+2+1=11$. We can now only 
subdivide by the remaining now-marginal operator $T_1$ since $T_8$ 
being irrelevant does not affect the conformal field theory. Realizing 
the lattice vector combinations \ 
${1\over 4} (23T_1-16\al_3-\al_1) = (0,1,0)$ \ and \ 
${1\over 4} (24T_2-2T_1-2\al_1) = (0,1,0)$, we see that the $T_1$ 
subdivision results in the subcones 
$C(0;T_1,\al_3,\al_1)$ and $C(0;T_2,T_1,\al_1)$, which are 
respectively $\BZ_4\ (1,0,1)$ and $\BZ_4\ (0,1,1)$ geometric terminal 
singularities (since the potential tachyons do not survive the Type II 
GSO projection). However we must realize that both of these are secretly 
supersymmetric $\BZ_4\ (1,-1)$ singularities when twisted states in 
the other (anti-)chiral rings are taken into account. Thus the final 
endpoint of the most-relevant-tachyon sequence in this Type II theory 
includes only flat and supersymmetric spaces. 

On the other hand, note that there are flip transitions (shown by the 
dotted lines) if $T_8$ condenses first followed by $T_2, T_1$, landing 
up at distinct endpoints via condensation of different sequences of 
tachyons. Since $T_2={1\over 4}(T_8+\al_2)$, $T_2$ remains relevant with 
renormalized R-charge ${1\over 2}$ after $T_8$ has condensed. The 
subsequent tachyon $T_1$ has renormalized R-charge 
\be
R_1'=\ell_{\Delta^{23T}}(T_1)={D^{23T}(T_1)\over D^{23T}(T_8)}=
{17\over 23}\ \biggl[\ 1 + {1\over 8}\ {1-{21\over 23}\over {17\over 23}}\ 
\biggr] = {3\over 4} > {17\over 23} 
\ee
The total volume of the subcones in this case is 
$V_{total}=9+6(1)=15>11$, which verifies the fact that the most 
relevant tachyon sequence gives minimal total volume for the subcones.

\section{Conclusions}

We have studied condensation of localized tachyons in $\BC^3/\BZ_N$
nonsupersymmetric orbifolds via the worldsheet RG flows induced
thereby.  We have seen that this generically leads to a set of
decoupled residual geometries that include $geometric$ terminal
singularities, with no marginal or relevant K\" ahler blowups by which
they can be resolved (although generic metric blowup modes generically
do exist). Treated as geometric spaces, they thus admit no canonical
resolution and the various possible distinct resolutions via
condensation of distinct sequences of tachyons are sometimes related
by flip transitions. In general, the renormalized R-charges of
subsequent tachyons in the residual geometries are higher than their
previous values. Thus the residual geometries in general are more
prone to becoming terminal singularities after tachyon condensation.
For Type II theories with no bulk tachyon, we have shown that all-ring
terminal singularities cannot exist, which shows that the endpoint of
tachyon condensation in Type II unstable $\BC^3/\BZ_N$ orbifold
theories are always smooth spaces. 

The calculations via toric geometry described in this paper are 
essentially a reflection of the physics underlying gauged linear sigma 
models. In particular, topological twisted GLSMs may be reliably used 
to study tachyon condensation not simply at the endpoints of but $all$ 
$along$ the worldsheet RG flow and map out the phase structure of two 
dimensional theories including tachyons. 

The methods we have used here are of course not powerful enough to study 
situations where, for instance, mixed tachyons (combinations of tachyons 
from distinct rings) condense simultaneously. In such cases, we lose 
control over the system because $(2,2)$ worldsheet supersymmetry breaks 
down. 

We now make a few brief comments on the physics seen by the worldvolume 
theory on a D-brane probe of a nonsupersymmetric orbifold. In general, 
closed string twist fields appear as Fayet-Iliopoulos D-term couplings 
in the D-brane probe theory \cite{douglasmoore, aps, hkmm}. 
Here, the closed string twist fields that are tachyonic condense in time 
and thus have a time-dependent expectation value, say of functional form
$T(t)$. Via the D-term equations, these induce time-dependent Higgs 
expectation values for the bifundamental link fields of the quiver, 
which are then proportional to $\sqrt{T(t)}$. For simplicity, let us 
assume that closed string tachyon condensation occurs so as to 
monotonically increase the condensate value $T(t)$. Then the link field 
expectation values also increase in time. Consider a low energy observer 
on a D-brane probe who observes physics at energy $E$. Then the link 
field vevs increase monotonically in time so that the link fields are 
naturally integrated out in time from the point of view of the low 
energy observer, thereby leaving a residual quiver with fewer link 
fields and a less singular \cite{aps} orbifold.\footnote{A (gauge-fixed) 
block-spin-like transformation that coarse-grains matrix 
representations of various D-brane configurations was studied in 
\cite{kn0211}. In particular \cite{knrp0309} studied (in part along 
similar lines) a block-spin-like transformation on a simplified subset 
of quiver gauge theories that arise on the worldvolumes of D-brane 
probes of supersymmetric orbifolds by sequentially Higgsing the gauge 
symmetry using the bifundamental scalar link fields present in these 
theories. From this point of view, the image branes for a 
nonsupersymmetric orbifold naturally form ``block-(image)branes'' in 
time in the process of condensation of a localized tachyon. For 
instance, as the link $X_{ij}$ is integrated out below energies $E$, 
the images $i$ and $j$ form the block-(image)brane $ij$. The 
``upstairs'' matrices of the image branes do not coarse-grain in the 
homogeneous fashion studied in \cite{kn0211, knrp0309}. Instead row 
$i$ and column $j$ are deleted from the $N\times N$ matrix to get the 
$(N-1)\times (N-1)$ ``upstairs'' matrices of the residual orbifold.}

It would be interesting to analyze the worldvolume D-brane gauge 
theories on $\BC^3/\BZ_N$ orbifold singularities and study their 
implications, in part with a view to constructing stable 
nonsupersymmetric string vacua.

\vspace{6mm}

%\newpage
{\small {\bf Acknowledgments:} It is a pleasure to thank P.~Argyres, 
J.~Maldacena, S.~Minwalla, J.~Polchinski and especially P.~Aspinwall and 
E.~Martinec for helpful discussions. KN thanks the organizers of the 
KITP Superstring Cosmology conference, Santa Barbara, USA and the IIT 
Kanpur String Workshop, Kanpur, India for hospitality during some 
stages of this work. 
This material is based upon work supported by the National Science 
Foundation under Grant Nos. PHY99-07949 and DMS-0074072.}

\appendix
\section{The GSO projection} 

Our discussion of the chiral GSO is based on and generalizes that 
appearing in \cite{atish94, lowestrom, aps, 
vafa0111, hkmm, emilrev, 0308029}. We first outline a 
method to engineer the GSO projection based essentially on its action 
on the untwisted sector and consistency thereof with supersymmetric 
orbifolds. Then we work out the RNS partition function and, by 
requiring modular invariance thereof, obtain the GSO projection. 
\\

{\bf \emph{``Engineering'' the GSO projection}}\\
Here we outline a method to engineer the chiral GSO projection for a Type 
II orbifold $\BC^3/\BZ_N\ (k_1,k_2,k_3)$ with $k_i$ not necessarily equal 
to one. We complexify the eight transverse untwisted fermions into 
$\psi_i=e^{iH_i},\ i=0,1,2,3$. Consider a symmetry acting on the untwisted 
(complex) fermions and the twist fields via $H_i \ra H_i + a_i \pi$, \ie\ 
\bea\label{gso1}
\psi_i &\ra& \psi_i\ {\rm e}^{ia_i\pi}, \nonumber\\
{} X_j &\ra& X_j\ \ {\rm exp} \biggl[i\pi \sum_i a_i \biggl\{ {jk_i\over N} 
\biggr\} \biggr]  \equiv X_j\ (-1)^{E_j} .
\eea
This defines a $(-1)^{F_L}$ $\BZ_2$ action on the untwisted sector thus 
eliminating the bulk tachyon only if the $a_i$ are odd integers. The 
action on the twisted states $X_j$ is a well-defined $\BZ_2$ if the 
exponent $E_j$ is an integer. This GSO exponent can be written as 
\be
E_j = \sum_i a_i \biggl\{ {jk_i\over N} \biggr\} 
= {j\over N}\ \sum_i a_i k_i - \sum_i a_i\biggl[ {jk_i\over N} \biggr] .
\ee
Thus $E_j$ is integral if we have $a_i=odd$ satisfying 
$\ \sum_i a_i k_i=0\ ({\rm mod\ 2N})$. Consider the case $\sum_i k_i=odd$. 
Then 
\be
\sum a_i k_i = a_1 k_1 + a_2 k_2 + a_3 ({\rm odd} - k_1 - k_2) 
= (a_1 - a_3) k_1 + (a_2 - a_3) k_2 + {\rm odd} = {\rm odd}
\ee
since the first two terms (containing differences of two odd integers) 
are each even. Thus this shows that no $a_i=odd$ exist satisfying 
$\ \sum_i a_i k_i=0\ ({\rm mod\ 2N})$: in other words, for even $N$ and 
$\sum_i k_i=odd$, no chiral GSO projection exists (note that for odd $N$, 
one can always shift say $k_3\ra k_3\pm N$ to make $\sum_i k_i$ even and 
thereby recover a chiral GSO projection). 

For $\sum_i k_i=even$,\ $E_j$ simplifies to 
\bea\label{gso2}
E_j = \sum_i a_i \biggl[ {jk_i\over N} \biggr] 
&=& \sum_i \biggl[ {jk_i\over N} \biggr] + {\rm even} \\ 
{} &=& {j\over N}\sum_i k_i - R_j + {\rm even}, \nonumber 
\eea
where $R_j=\sum_i \{ {jk_i\over N} \}$ is the R-charge of the twisted 
state. Thus for a twisted state $T$ with R-charge $R=(r_1,r_2,r_3)$ in 
the orbifold $\BC^3/\BZ_N\ (k_1,k_2,k_3)$, the GSO exponent is 
\be\label{gsorpq}
E = \sum_i a_i r_i, \qquad {\rm with}\ \ 
a_i={\rm odd\ and}\ \sum a_i k_i = 0\ ({\rm mod} 2N) .
\ee

Let us examine this GSO exponent in greater detail to elucidate its 
properties.

For $\BC/\BZ_N$, consider the action $H\ra H+a\pi$, \ie\ 
\be
\psi \ra \psi\ (-1)^a, \qquad \qquad X_j \ra X_j\ (-1)^{ja/N}, 
\ee
which defines a nontrivial $(-1)^{F_L}$ $\BZ_2$ action on the 
untwisted sector only if $a$ is an odd integer. For odd $N=2M+1$ prime 
and general odd $a=2b+1$, the action on the twisted states $X_j$ defines 
a $\BZ_2$ only if $a=N$. For other odd $N$, there are appropriate 
twist-$j$ subsectors with prime factors for which the argument is then 
the same. Indeed let us consider even $N=2^k$ with $a=2b+1$, so that the 
twisted states transform as 
\be
X_j \ra X_j\ (-1)^{ja/N} = X_j\ (-1)^{j(2b+1)/(2^k)}
\ee
Then for no twist-$j$ subsector is there a well-defined $\BZ_2$. For 
general even N=2M, there are twist-j subsectors which are identical to 
the above cases. Thus we take $a=N$ with $N$ odd for Type II $\BC/\BZ_N$. 
Then the chiral GSO above acts as 
\be
\psi \ra \psi\ (-1)^N, \qquad \qquad X_j \ra X_j\ (-1)^j .
\ee
Since the $(-1,-1)$ picture vertex operators are odd under chiral GSO, 
only the $X_j$ states with odd $j$ survive. 

Now consider $\BC^2/\BZ_{N(p)}$ and $\BC^3/\BZ_{N(p,q)}$. In these cases, 
the GSO exponent 
$E_j$ is an integer if $\ a_1+a_2p+a_3q=0({\rm mod\ 2N})$.

From above, we have $ \ 1+p+q=even$. 
For $\BC^2/\BZ_{N(p)}$, we have $q=0$: then $p=odd$ and $a_1=p, a_2=-1$ 
satisfy $\ a_1+a_2p=0\ ({\rm mod} 2N)$. We have then the chiral GSO action 
\be\label{gsoC2} 
\psi_1 \ra \psi_1\ (-1)^p, \ \ \psi_2 \ra \psi_2\ (-1), \qquad \qquad 
X_j \ra X_j\ (-1)^{[jp/N]}
\ee
for the $\BC^2/\BZ_{N(p)}$ orbifold. 

Note that the supersymmetric orbifold is $p=-1$ in this convention, so 
that $[{jp\over N}]=[-{j\over N}]=-1$. Thus the chiral GSO acts as 
$X_j\ra X_j (-1)$ for all twisted states, which in fact are marginal 
with R-charge $R_j={j\over N}+ \{ -{j\over N} \}=1$. Thus the entire 
$(c_X,c_Y)$ ring is preserved by the chiral GSO in this case, while the 
entire $(c_X,a_Y)$ ring is projected out. 

For $\BC^3/\BZ_{N(p,q)}$, we have $p+q=odd$: then $a_1=p+q, a_2=a_3=-1$ 
satisfy $\ a_1+a_2p+a_3q=0({\rm mod\ 2N})$. Then the GSO exponent $E_j$ is 
\be
E_j = \biggl[ {jp\over N} \biggr] + \biggl[ {jq\over N} \biggr] 
= {j\over N}\ (1+p+q) - {j\over N} - \biggl\{ {jp \over N} \biggr\} 
- \biggl\{ {jq \over N} \biggr\} = {j\over N}\ (1+p+q) - R_j .
\ee
and we have the chiral GSO action 
\be\label{gsoC3} 
\psi_1 \ra \psi_1\ (-1)^{p+q}, \ \ \psi_2 \ra \psi_2\ (-1), 
\ \ \psi_3 \ra \psi_3\ (-1), \qquad 
X_j \ra X_j\ (-1)^{[jp/N]+[jq/N]} .
\ee
For the supersymmetric case $1+p+q=0$ and we have $\ E=-R_j\ $, so that 
the chiral GSO acts as $X_j\ra X_j (-1)$ for all twisted states in the 
$(c_X,c_Y,c_Z)$ ring with R-charge $R_j=1$. The blowup modes of the 
geometry are defined purely in terms of the marginal $R_j=1$ states for 
the supersymmetric case so that they are all preserved by the GSO (note 
that there also exist $R_j=2$ twisted states here, unlike in the 
$\BC^2/\BZ_{N(p)}$ case). Thus in general, the chiral GSO for Type II 
theories projects out the twist operators $X_j$ of the $(c_X,c_Y,c_Z)$ 
ring with $[{jp\over N}]+[{jq\over N}]\in 2\BZ$ and retains those 
with either $[{jp\over N}]$ or $[{jq\over N}]$ odd (but not both). 
Similarly in the $(c_X,c_Y,a_Z)$ ring, the $X_j$ with 
$[{jp\over N}]+[{jq\over N}]$ even are retained and so on. 

Note that this agrees with the Green-Schwarz GSO analysis (generalized 
from that in \cite{aps}), which starts with a rotation generator 
\be
R={\rm exp}\biggl[{2\pi i\over N} (J_{45} + p J_{67} + q J_{89}) \biggr]
\ee
Then 
\be
R^N = (-1)^{2(s_{45} + ps_{67} + qs_{89})}
\ee
so that considering the action thereof on various spinor charge sectors
$(\pm 1/2,\pm 1/2,\pm 1/2)$, and demanding $R^N=1$ for removing the bulk
tachyon gives $p+q=odd$. 

It is important to note that changing $q\ra q+N$ introduces an extra
factor of $(-1)^N$ in (\ref{gsoC3}). Also
$p+q=odd$ now changes to $p+q+N=even$ thus reversing the bulk tachyon
projection and changing a Type II theory to Type 0.  For example, the
$\BC^3/\BZ_{11(2,-7)}$ orbifold is a good Type II theory with the bulk
tachyon projected out, whereas the orbifold $\BC^3/\BZ_{11(2,4)}$
(which is equivalent in conformal field theory) does have a bulk
tachyon and should be regarded as Type 0.  Furthermore, the action on
the twisted states changes since the exponent $E$ becomes $E\ra j+E$,
so that supersymmetric orbifolds now have 
$E\ra {j\over N}(1+p+q+N)-R_j=j-R_j$, thus projecting down to 
$j=even$ twisted states among the $R_j=1$ blowup modes.
\\

{\bf \emph{Modular invariance of the partition function}} \\
The partition function for $\BC/\BZ_N$ was described notably in 
\cite{atish94, lowestrom}, using both the Green-Schwarz and the 
RNS formulations. We primarily use the RNS formulation here.

The Type 0 string has a diagonal GSO projection that ties together the 
left and right movers: it has the partition function 
\bea
Z &=& {1\over 2N} \sum_{j,l=0}^{N-1}\ 
{1\over |\eta^2(\tau) {\bar \eta}^2({\bar \tau})|}\ 
\prod_{i=1}^3\ \Biggl| {\eta(\tau)\over \theta 
[ {%\footnotesize 
\bA{c} {1\over 2}+jk_i/N \\ 
{1\over 2}+lk_i/N \\ \eA }](0,\tau)} \Biggr|^2 \cdot \qquad \qquad 
\nonumber\\ 
\qquad \qquad && {} \Biggl[ \ \ \biggl| \prod_{i=1}^3 
\theta \left[ {%\small 
\bA{c} {jk_i\over N} \\ {lk_i\over N} \\ \eA 
}\right]\ \theta \left[ {%\small 
\bA{c} 0 \\ 0 \\ \eA }\right] \biggr|^2 
+ 
\biggl| \prod_{i=1}^3 \theta \left[ {%\small 
\bA{c} {jk_i\over N} \\ 
{lk_i\over N} + {1\over 2} \\ \eA }\right]\ 
\theta \left[ {%\small 
\bA{c} 0 \\ {1\over 2} \\ \eA }\right] \biggr|^2 
\\
\ \qquad \qquad && {} + \biggl| \prod_{i=1}^3 
\theta \left[ {%\small 
\bA{c} {jk_i\over N} + {1\over 2} 
\\{lk_i\over N} \\ \eA }\right]\ 
\theta \left[ {%\small 
\bA{c} {1\over 2} \\ 0 \\ \eA }\right] \biggr|^2 
\pm \biggl| 
\prod_{i=1}^3 \theta \left[ {%\small 
\bA{c} {jk_i\over N} + {1\over 2} 
\\ {lk_i\over N} + {1\over 2} \\ \eA }\right]\ 
\theta \left[ {%\small 
\bA{c} {1\over 2} \\ {1\over 2} \\ \eA }\right]
\biggr|^2 \ \ \Biggr] \nonumber
\eea
Clearly this partition function exists for any $k_i,N$. 

On the other hand, the 1-loop partition function on a 
$\BC^3/\BZ_N\ (k_1,k_2,k_3)$ orbifold for a Type II string with separate 
GSO projections on the left and right movers is given by the sum over 
twisted sectors as 
\be
Z = {1\over 4N} \sum_{j,l=0}^{N-1}\ 
{1\over |\eta^2(\tau) {\bar \eta}^2({\bar \tau})|}\ 
\prod_{i=1}^3\ \Biggl| {\eta(\tau)\over \theta 
[ {%\footnotesize 
\bA{c} {1\over 2}+jk_i/N \\ 
{1\over 2}+lk_i/N \\ \eA }](0,\tau)} \Biggr|^2 \ 
\biggl| {\zeta^j_l \over |\eta^4(\tau)} \biggr|^2
\ee
where 
\bea
\zeta^j_l &=& \prod_{i=1}^3 
\theta \left[ {%\small 
\bA{c} {jk_i\over N} \\ {lk_i\over N} 
\\ \eA }\right]\ \theta \left[ {%\small 
\bA{c} 0 \\ 0 \\ \eA }\right] 
- e^{-i\pi \sum_i {jk_i\over N}} 
\prod_{i=1}^3 \theta \left[ {%\small 
\bA{c} {jk_i\over N} \\ 
{lk_i\over N} + {1\over 2} \\ \eA }\right]\ 
\theta \left[ {%\small 
\bA{c} 0 \\ {1\over 2} \\ \eA }\right] \nonumber\\
\ \ && {} - \prod_{i=1}^3 
\theta \left[ {%\small 
\bA{c} {jk_i\over N} + {1\over 2} 
\\{lk_i\over N} \\ \eA }\right]\ 
\theta \left[ {%\small 
\bA{c} {1\over 2} \\ 0 \\ \eA }\right] 
- e^{-i\pi \sum_i {jk_i\over N} } 
\prod_{i=1}^3 \theta \left[ {%\small 
\bA{c} {jk_i\over N} + {1\over 2} 
\\ {lk_i\over N} + {1\over 2} \\ \eA }\right]\ 
\theta \left[ {%\small 
\bA{c} {1\over 2} \\ {1\over 2} \\ \eA }\right] 
\eea
contains the sum over spin structures for the j-th twisted sector 
twisted by $g^l$ in the ``time'' direction. The terms in Z are easily 
recognized as the contributions from the untwisted bosons in the one 
complex flat dimension, multiplied by the contributions from the twisted 
bosons in the three orbifolded complex dimensions and the fermionic 
contributions. 

At this point, we list some formulae involving theta functions that we 
use here (from \cite{atish94, joetext}). The boundary conditions 
on the worldsheet scalars and spinors in the NS sector are 
\bea
X(w+2\pi) &=& e^{2\pi i a} X(w), \qquad X(w+2\pi \tau) = e^{2\pi i b} X(w), 
\nonumber\\
\psi(w+2\pi) &=& -e^{2\pi i a} \psi(w) \equiv - e^{-i\pi \al} \psi(w), 
\nonumber\\
\psi(w+2\pi \tau) &=& -e^{2\pi i b} \psi(w) \equiv - e^{-i\pi \beta} \psi(w) .
\eea
A chiral fermion with Hamiltonian 
\be
H_a = \sum_{n=1}^{\infty} \biggl[
(n - {1\over 2} + a) d^{\dag}_n d_n + 
(n - {1\over 2} - a) {\bar d}^{\dag}_n {\bar d}_n \biggr] + 
{a^2\over 2} - {1\over 24}
\ee
has the partition function 
\bea
Z^{\al}_{\beta} &=& Tr\ (h_b q^{H_a}) = 
{\theta[{%\footnotesize 
\bA{c} a \\ b \\ \eA }] \over \eta(\tau)} \equiv 
{\theta[{%\footnotesize 
\bA{c} -{\al\over 2} \\ -{\beta\over 2} \\ \eA }] 
\over \eta(\tau)} \nonumber\\
&=& e^{2\pi iab}\ q^{{a^2\over 2} - {1\over 24}}\ 
\prod_{n=1}^{\infty} (1 + q^{n - {1\over 2} + a}\ e^{2\pi i b})
(1 + q^{n - {1\over 2} - a}\ e^{-2\pi i b})
\eea
where the $\BZ_N$ action on the Hilbert space is 
\be
h_b d h_b^{-1} = -e^{-2\pi i b} d, \qquad 
{\bar h}_b {\bar d} {\bar h}_b^{-1} = -e^{2\pi i b} {\bar d}
\ee
Some useful formulae involving theta functions are
\bea
&& \theta[{%\footnotesize 
\bA{c} a \\ b \\ \eA }](\nu,\tau) = 
e^{i\pi a^2\tau + 2\pi ia(\nu + b)}\ 
\theta[{%\footnotesize 
\bA{c} 0 \\ 0 \\ \eA }](\nu + a\tau + b,\tau) 
= \sum_{n\in \BZ} q^{{1\over 2}(n+a)^2}\ e^{2\pi i (n+a)(\nu + b)}
\nonumber\\
&& {} \theta[{%\footnotesize 
\bA{c} a+r \\ b+s \\ \eA }](\nu,\tau) = 
e^{i\pi isa} \theta[{%\footnotesize 
\bA{c} a \\ b \\ \eA }](\nu,\tau), 
\qquad r,s\in \BZ
\eea
Under modular transformations generated by $S:\tau \ra -{1\over \tau}$ 
and $T:\tau \ra \tau + 1$, we have 
\bea
&& T: \ \theta[{%\footnotesize 
\bA{c} a \\ b \\ \eA }] \ra 
e^{-i\pi a^2 - i\pi a}\ 
\theta[{%\footnotesize 
\bA{c} a \\ a+b+{1\over 2} \\ \eA }], 
\qquad \eta \ra e^{i\pi/12}\ \eta 
\nonumber\\ 
&& {} S: \ \theta[{%\footnotesize 
\bA{c} a \\ b \\ \eA }] \ra 
(-i\tau)^{1\over 2} e^{2\pi iab}\ 
\theta[{%\footnotesize 
\bA{c} -b \\ a \\ \eA }], 
\qquad \eta \ra (-i\tau)^{1\over 2} \eta
\eea

Thus under an S-transformation, we have 
\bea
{\zeta^j_l\over \eta^4} &\ra& e^{2\pi i\sum_i {jl k_i^2\over N^2}} 
\Biggl\{ \prod_{i=1}^3 
\theta\left[{%\small 
\bA{c} -{lk_i\over N} \\ {jk_i\over N} 
\\ \eA }\right]\ \theta\left[{%\small 
\bA{c} 0 \\ 0 \\ \eA }\right] 
- e^{-i\pi \sum_i {jk_i\over N} + 2\pi i\sum_i {1\over 2} {jk_i\over N}} 
\prod_{i=1}^3 \theta\left[{%\small 
\bA{c} -{lk_i\over N}-{1\over 2} \\ 
{jk_i\over N} \\ \eA }\right] 
\theta\left[{%\small 
\bA{c} -{1\over 2} \\ 0 \\ \eA }\right] 
\nonumber\\ \ \ && {} 
- e^{2\pi i\sum_i {1\over 2} {lk_i\over N}} \prod_{i=1}^3 
\theta\left[{%\small 
\bA{c} -{lk_i\over N} 
\\{jk_i\over N}+{1\over 2} \\ \eA }\right]\ 
\theta\left[{%\small 
\bA{c} 0 \\ {1\over 2} \\ \eA }\right] \nonumber\\
\ \ && {} 
- e^{-i\pi \sum_i {jk_i\over N} + 2\pi i\sum_i {1\over 2} ({jk_i\over N} 
+ {lk_i\over N}) + 2\pi i({3\over 4}+{1\over 4})} 
\prod_{i=1}^3 \theta\left[{%\small 
\bA{c} -{lk_i\over N} - {1\over 2} 
\\ {jk_i\over N} + {1\over 2} \\ \eA }\right]\ 
\theta\left[{%\small 
\bA{c} -{1\over 2} \\ {1\over 2} \\ \eA }\right] 
\Biggr\}\ .
\eea
The sum over such terms can be rewritten as the original partition 
function with $j'=N-l,\ l'=j$ if the phase from the third term above 
satisfies 
\be
e^{i\pi \sum_i {lk_i\over N} } = e^{-i\pi \sum_i {j'k_i\over N} } 
= e^{-i\pi \sum_i {(N-l)k_i\over N} } 
\ee
In other words, we require 
\be
\sum_i {(N-l)k_i\over N} = -\sum_i {lk_i\over N} + {\rm even}
\ee
\ie\ $\sum_i k_i=$ even, the condition we have seen before (this 
condition on the orbifold weights can also be obtained by demanding 
level-matching). Invariance under the T transformation does not give 
anything new. 

We can now expand the Type II partition function we have here to realize 
the GSO projection on the twisted states to obtain the projector 
\be\label{projccc}
1 - (-1)^{-i\pi \sum_i [jk_i/N]}
\ee
for the ground states in the sector where $\{ {jk_i\over N} \} < 
{1\over 2}$, \ie\ the $(c_X,c_Y,c_Z)$ ring. This is a projector onto 
twisted states with \ $\sum_i [jk_i/N]=E_j=odd$, recovering the result 
from the previous subsection. On the other hand, consider as an example, 
the sector where $\{ {jk_3\over N} \} > {1\over 2}$ with 
$\{ {jk_1\over N} \}, \{ {jk_2\over N} \} < {1\over 2}$. Then we obtain 
the projector 
\be\label{projcca}
1 - (-1)^{-i\pi (\sum_i [jk_i/N] - 1)}
\ee
for the ground states (which are in the $(c_X,c_Y,a_Z)$ ring), \ie\ 
$\sum_i [jk_i/N]=even$. The chiral operators $X_j$ are obtained as the 
excited state with one extra fermion number from $\psi_3$ which therefore 
have the GSO projection $\sum_i [jk_i/N]=E_j=odd$, as before. Likewise if 
two of $\{ {jk_i\over N} \} > {1\over 2}$, we have $E_j=odd$ for the ground 
states so that the $X_j$, obtained with one extra fermion number in the 
two sectors, again have $E_j=odd$ and so on. Thus the GSO exponent for 
the chiral operators $X_j$ is $E_j=\sum_i [jk_i/N]=odd$. 

The above partition function can be recast as the Green-Schwarz partition 
function using the quartic Riemann identity for theta functions 
\cite{0308029} (see \cite{atish94, lowestrom} for the $\BC/\BZ_N$ 
case). It is noteworthy that the partition function of the Type II theory 
can be obtained by gauging a chiral $(-1)^{F_L}$ $\BZ_2$ symmetry in 
(\ie\ as a $\BZ_2$ orbifold of) the partition function of the Type 0 
theory and demanding modular invariance, along the lines of \cite{sw86,
amv86}. In this case, this procedure effectively changes the $\BZ_N$ 
orbifold to a $\BZ_N\times \BZ_2$. For odd $N$, this is the same as 
$\BZ_{2N}$ and we recover the partition function above and thence the 
Green-Schwarz partition function. While in principle one could expect 
interesting generalizations involving discrete torsion in 
$\BZ_N\times \BZ_2$ for even $N$, a careful calculation shows that 
the possible extra phases can be absorbed via redefinitions of the 
orbifold weights.

\section{$\BC^2/\BZ_{N(p)}$ toric geometry}

We outline here the toric description of $\BC^2/\BZ_{N(p)}$ discussed 
in \cite{hkmm} (see also \cite{emilrev}), based on the Hirzebruch-Jung 
theory of singularity resolution in codimension two. We have 
uniformized our notations and conventions with our description of 
$\BC^3/\BZ_{N(p,q)}$.

A basis for monomials invariant under the orbifold action is 
$\ u=x^N,\ v=x^{-p}y$,\ \ so that the ring of holomorphic functions on 
a neighbourhood of the noncompact $\BC^2/\BZ_N$ singularity is generated 
by the monomials $\ u^{m_1} v^{m_2}=x^{Nm_1-pm_2}\ y^{m_2}\ $ for 
integer $m_1,m_2$, which gives a cone in the $\BM$ lattice bounded by 
$(1,0),\ (p,N)\ $. From these, we read off the vertices of the fan in 
the $\BN$ lattice dual to this 
\be
\al_1=(N,-p),\qquad \al_2=(0,1) .
\ee
This uniformizes our notations and conventions with those of \cite{hkmm}:
see \eg\ Figure~2 therein (see also \eg\ Figure~1.10 of \cite{emilrev}). 
The Hirzebruch-Jung formulation of minimal resolution in codimension two 
ensures that the endpoint of condensation of all tachyons in a given 
orbifold is always flat space. Thus the volume of any subcone arising 
from a subdivision by a tachyonic blowup mode is equal to one, so that 
unlike $\BC^3/\BZ_{N(p,q)}$, terminal singularities do not exist here. 

Furthermore there is a nice continued fraction representation of 
${N\over p}$ for $\BC^2/\BZ_{N(p)}$, in terms of integers $a_k$, 
following from the Hirzebruch-Jung theory of minimal resolution 
of these orbifolds which encodes the relations 
\be
a_j v_j = v_{j-1} + v_{j+1}
\ee
between the vectors $v_0=(N,-p),\ v_{r+1}=(0,1),\ v_j,\ j=1,\ldots,r$ 
in the toric diagram representing the bounding vectors of the fan as 
well as the $r$ generators of the chiral ring. Then by iterating the 
relation above, we can see from the toric diagram that $v_k$ obeys 
\be
A_k v_k = B_k v_0 + v_{k+1},
\ee
where
\be
A_k = a_k - 1/(a_{k-1} - ...),\ \  A_1=a_1, \ \qquad 
B_k = [ A_{k-1}. A_{k-2} ... A_1 ]^{-1} .
\ee
Then, realizing that (1,0) is an interior point for $p<N$, we have for 
$v_r = (1,0)$,
\be
A_r (1,0) = B_r (N,-p) + (0,1)
\ee
so that $\ A_r=B_r N,\ 1=B_r p$,\ \ giving 
\be
{N\over p}=A_r = [a_r,a_{r-1} \ldots ] .
\ee
There does not appear to be any generalization of the continued fraction 
fraction representation for $\BC^3/\BZ_{N(p,q)}$. 

In this case also, one can study subsequent tachyons Type II theories in 
light of the chiral GSO projection. For a $\BC^2/\BZ_{N(p)}$ orbifold 
that is Type II, we have $p=odd$. A twisted state 
$T_j=(j,-[{jp\over N}])\equiv ({j\over N}, \{ {jp\over N} \})$ is 
preserved by the GSO if $[{jp\over N}]=odd$. Consider blowing up the 
divisor corresponding to $T_j$. The subcone $C(0;T_j,\al_2)$ can be 
easily seen to be a $\BC^2/\BZ_j\ (1,[{jp\over N}])$ orbifold which 
clearly admits a Type II projection since $1+[{jp\over N}]=even$. If 
$T_k\in C(0;T_j,\al_2)$, solving $T_k=aT_j+b\al_2$ gives 
\be
R_k' = (a,b) = \biggl({k\over j},\ \ {k\over j} \biggl[{jp\over N}\biggr] 
- \biggl[{kp\over N}\biggr]\biggr) .
\ee
The GSO requires $a_i=odd$ satisfying 
$a_1+a_2[{jp\over N}]=0\ ({\rm mod} 2j)$, which is solved by 
$a_1=[{jp\over N}], a_2=-1$. Thus the GSO exponent for $T_k$ works out to 
\be
E = \biggl[{jp\over N}\biggr] {k\over j} - 
{k\over j} \biggl[{jp\over N}\biggr] + \biggl[{kp\over N}\biggr] 
= \biggl[{kp\over N}\biggr] .
\ee
Similarly, the subcone $C(0;T_j,\al_1)$ is seen to be a 
$\BC^2/\BZ_{N\{ {jp\over N} \} }\ (p,-[{jp\over N}])$ orbifold (assuming 
nondegeneracy of the Smith normal form vector). This also clearly admits 
a Type II projection since $p-[{jp\over N}]=even$. A sub-twisted state 
$T_k\in C(0;T_j,\al_1)$ satisfies $T_k=aT_j+b\al_1$ giving
\be
R_k' = (a,b) = \biggl( { \{ {kp\over N} \}\over \{ {jp\over N} \} }, 
\ \ {k\over N} - {j\over N} { \{ {kp\over N} \}\over \{ {jp\over N} \}}
\biggr) .
\ee
A set of $a_i=odd$ satisfying 
$a_1p-a_2[{jp\over N}]=0\ ({\rm mod} 2N\{ {jp\over N} \})$, is 
$a_1=[{jp\over N}], a_2=p$, giving for the GSO exponent for $T_k$ 
\be
E = \biggl[{kp\over N}\biggr] .
\ee
Thus originally preserved sub-twisted states remain preserved after a 
preserved twisted state condenses. The R-charges of subsequent tachyons 
again expectedly renormalize upwards, \ie\ $R_k'> R_k$, since a sub-tachyon 
is closer to a sub-plane than to the original plane of marginal operators. 
However due to an interesting convexity property of the tachyonic 
generators (that appear in the Hirzebruch-Jung continued fraction) inside 
a given cone of a toric digram (see \eg\ \cite{martinecmoore}), it is 
straightforward to show that $R_k'\leq 1$ always! Thus there are no 
geometric terminal singularities in codimension two.

\end{document}